\documentclass[aps,prd,twocolumn,preprintnumbers,superscriptaddress,nofootinbib,longbibliography,floatfix]{revtex4-1}

\usepackage{amsmath}
\usepackage{graphicx}
\usepackage{dcolumn}
\usepackage{bm}
\usepackage[caption=false]{subfig}
\usepackage{xspace}
\usepackage[countmax]{subfloat}
\usepackage{color}
\definecolor{darkblue}{rgb}{0,0,0.5}

\newcommand{\aNLO}[1]{{\textsc{MadGraph5\_aMC@NLO\xspace #1}}}
\newcommand{\fastjet}[1]{\textsc{FastJet\xspace #1}}
\newcommand{\herwig}[1]{\textsc{Herwig\xspace #1}}
\newcommand{\herwigpp}[1]{\textsc{Herwig\xspace #1}}
\newcommand{\pythia}[1]{\textsc{Pythia\xspace #1}}
\newcommand{\sherpa}[1]{\textsc{Sherpa\xspace #1}}

\newcommand{\Fig}[1]{Fig.~\ref{#1}}
\newcommand{\Figs}[2]{Figs.~\ref{#1} and \ref{#2}}

\newcommand{\App}[1]{App.~\ref{#1}}
\newcommand{\Sec}[1]{Sec.~\ref{#1}}
\newcommand{\Secs}[2]{Secs.~\ref{#1} and \ref{#2}}

\newcommand{\Ref}[1]{Ref.~\cite{#1}}
\newcommand{\Refs}[1]{Refs.~\cite{#1}}

\newcommand{\Eq}[1]{Eq.~(\ref{#1})}
\newcommand{\be}{\begin{equation}}
\newcommand{\ee}{\end{equation}}
\newcommand{\GeV}{\text{GeV}}
\newcommand{\TeV}{\text{TeV}}
\newcommand{\pb}{\text{pb}}

\newcommand{\df}{\text{d}}

\begin{document}

\title{Resurrecting the Dead Cone}

\preprint{\hbox{CP3--16--27, MIT--CTP 4810}}

\author{Fabio Maltoni}
\email{fabio.maltoni@uclouvain.be} \affiliation{Centre for
Cosmology, Particle Physics and Phenomenology CP3, Universit\'e
Catholique de Louvain, Chemin du Cyclotron, 1348 Louvain la
Neuve, Belgium}
\author{Michele Selvaggi}
\email{michele.selvaggi@uclouvain.be} \affiliation{Centre for
Cosmology, Particle Physics and Phenomenology CP3, Universit\'e
Catholique de Louvain, Chemin du Cyclotron, 1348 Louvain la
Neuve, Belgium}
\author{Jesse Thaler}
\email{jthaler@mit.edu} \affiliation{Center for Theoretical Physics, Massachusetts Institute of Technology, Cambridge, MA 02139, USA}


\begin{abstract}
The dead cone is a well-known effect in gauge theories, where radiation from a charged particle of mass $m$ and energy $E$ is suppressed within an angular size of $m/E$.
This effect is universal as it does not depend on the spin of the particle nor on the nature of the gauge interaction.
It is challenging to directly measure the dead cone at colliders, however, since the region of suppressed radiation either is too small to be resolved or is filled by the decay products of the massive particle.
In this paper, we propose to use jet substructure techniques to expose the dead cone effect in the strong-force radiation pattern around boosted top quarks at the Large Hadron Collider.
Our study shows that with 300/fb of 13-14 TeV collision data, ATLAS and CMS could obtain the first direct evidence of the dead cone effect and test its basic features.

\end{abstract}


\maketitle


\section{Introduction}

When charged particles are produced in high-energy collisions, they are usually accompanied by final-state radiation (FSR).  This process is familiar in quantum electrodynamics, where electrons radiate photons, as well as in quantum chromodynamics (QCD), where quarks (and gluons) radiate gluons.  The pattern of radiation depends crucially on the mass of the emitter but not on its spin, leading to the famous \emph{dead cone} effect \cite{Dokshitzer:1991fc,Dokshitzer:1991fd,Ellis:1991qj} where radiation from quarks with mass $m_q$ and energy $E_q$ is suppressed for emission angles $\theta \lesssim m_q/E_q$.  The dead cone is a fundamental prediction of QCD and other gauge theories, relying on only the behavior of radiation from massive particles in the soft (and collinear) limit.

While the prediction of the dead cone effect is uncontroversial, actually measuring the dead cone radiation pattern in QCD has turned out to be extremely challenging.  The reason is simple: massive particles decay, and the same angular scale $m/E$ appears both in the dead cone effect as well as in the characteristic opening angle between the decay products.  In this way, the dead cone is effectively ``filled'', so while the overall suppression of gluon radiation for heavy quarks can be inferred through inclusive~\cite{Schumm:1992xt,Dokshitzer:1995ev,Dokshitzer:2005ri,Perieanu:2006vn,Abdallah:2008ac} or semi-inclusive~\cite{Marchesini:1989yk,Okabe:1997uf,Ramos:2010cma} observables, the universal angular radiation pattern around the massive particle is obscured.  More direct probes of the dead cone for bottom and charm quarks have been put forward, for example in $e^+ e^-$ collisions at LEP \cite{Barlow:1991fe,Battaglia:989441} and in $ep$ collisions at HERA \cite{Chekanov:2000et}, and the dead cone effect is included in the shower deconstruction approach to top tagging \cite{Soper:2012pb}. To our knowledge, though, no definitive dead cone measurement has been made to date.

In this paper, we propose to directly measure the dead cone around top quarks at the Large Hadron Collider (LHC) using jet substructure techniques \cite{Abdesselam:2010pt, Altheimer:2012mn, Altheimer:2013yza, Adams:2015hiv}.  This direct approach is interesting both as a fundamental test of gauge theories and as a way to validate the treatment of radiation from massive particles in Monte Carlo generators.  Our focus is on leptonic decays, where a high-energy top quark can emit FSR gluons before decaying to a charged lepton, neutrino, and bottom quark.  Because $m_b/E_b \ll 1$ in top decays, the dead cone effect for the bottom quark is negligible, yet bottom-quark FSR is abundant and it tends to fill the top-quark dead cone region.  Using recursive jet clustering algorithms, though, we show how to statistically separate radiation from the top and bottom, thereby revealing the top-quark dead cone pattern.

Our method relies on soft drop declustering \cite{Larkoski:2014wba} (see also \Refs{Butterworth:2008iy,Krohn:2009th,Ellis:2009su,Ellis:2009me,Dasgupta:2013ihk}), a jet substructure technique that removes soft radiation from a jet to identify the hard jet core.  In \Ref{Larkoski:2015lea}, soft drop was applied to light quark and gluon jets to expose the famous Altarelli-Parisi splitting functions \cite{Altarelli:1977zs} which encode the energy pattern of FSR.  Here, we apply a similar technique to boosted top-quark jets, focusing now on the angular pattern of FSR.  While simple in its essence, our method relies on several key steps, such as the reconstruction of the top momentum despite the lost neutrino, whose robustness we test using parton shower (PS) generators.

The rest of this paper is organized as follows.  In \Sec{sec:leptoncase}, we review the dead cone effect in the idealized context of stable top quarks in electron-positron collisions.  In \Sec{sec:width}, we discuss subtleties related to top decay, and the contamination coming from initial state radiation (ISR) and underlying event (UE).   We present our novel measurement strategy in \Sec{sec:strategy} and estimate the LHC sensitivity with $300~\text{fb}^{-1}$ in \Sec{sec:reach}.  We discuss subdominant backgrounds in \Sec{sec:backgrounds} and conclude in \Sec{sec:conclude}, leaving additional cross checks to the appendices.


\section{Idealized top dead cone}
\label{sec:leptoncase}

We begin with the idealized case of top pair production in electron-positron collisions, $e^+ e^- \to t \bar{t}$, where we treat the top quark as stable. This approximation allows us to study the pattern of QCD radiation from the top quarks at various levels of accuracy without having to consider the top-quark decay products.

At tree level, each top quark carries three-momentum ${\rm p}_{t} = \sqrt{E_{t}^2-m_{t}^2}$ and energy $E_{t}=\sqrt{s}/2$, where  $\sqrt{s}$ is the $e^+ e^-$ collision energy.\footnote{Throughout this paper, the ``$t$'' subscript indicates the top quark, not to be confused with $p_T$ indicating transverse momentum with respect to the beam line in LHC collisions.}  In the soft and collinear limit, the probability for a top quark to emit an FSR gluon with energy fraction $z$ and opening angle $\theta$ is~\cite{Dokshitzer:1991fc,Dokshitzer:1991fd,Ellis:1991qj}
\begin{equation}
\frac{1}{\sigma}\frac{\df^2 \sigma}{\df z \, \df \theta^2} \simeq \frac{\alpha_S}{\pi} C_F \frac{1}{z}\frac{\theta^2}{(\theta^2 + \theta_{D}^2)^2},
\label{eq:dc}
\end{equation}
where $\alpha_s$ is the strong coupling constant, $C_F = 4/3$ is the top-quark color factor, $m_t \simeq 173~\GeV$ is the top mass, and
\be
\theta_{D} \equiv \frac{m_{t}}{{\rm p}_{t}} \simeq \frac{m_t}{E_t}
\ee
is the dead cone angle.  The relation $\theta_{D} \simeq m_{t}/E_{t}$ is valid already for moderately relativistic tops (e.g.~${\rm p}_t \gtrsim 2 m_t$), so we use this approximation throughout.  The emission probability in \Eq{eq:dc} reaches its maximum at $\theta \simeq \theta_{D}$ and is suppressed for angles $\theta \lesssim \theta_{D}$ (i.e.~a dead cone).   In the large Lorentz boost limit $m_t/E_t  \ll 1$, one recovers the usual collinear divergence for gluon emissions from massless quarks.  It is a remarkable property of the soft limit that the angular dependence in \Eq{eq:dc} is universal and does not depend on the spin of the massive particle.

\begin{figure}[t]
\includegraphics[width=0.95\columnwidth]{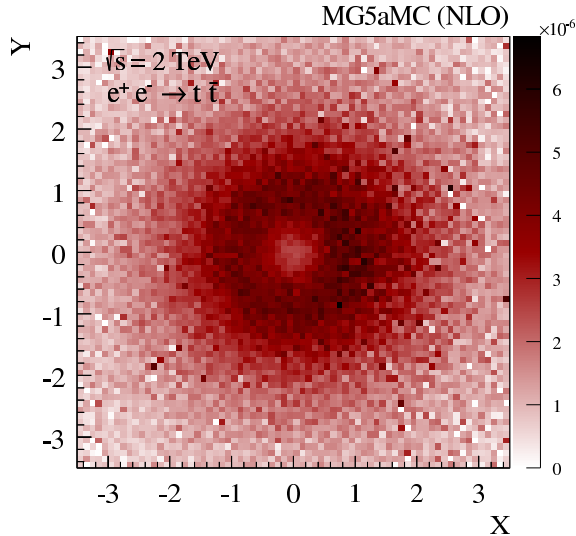}
\caption{Idealized gluon radiation pattern for $e^+ e^- \to t \bar{t}$ at $\sqrt{s} = 2~\TeV$, showing the expected dead cone suppression at the origin.  This is an NLO calculation with up to two additional partons in the final state.  To define the effective $t^* \to t g$ kinematics, the ``gluon'' corresponds to the sum of emissions within the top hemisphere, imposing a cut of $E_g > 50~\GeV$.  The $X$ and $Y$ coordinates are then normalized such that the dead cone peak is at $X^2 + Y^2 \simeq 1$.}
\label{fig:ee_tt_2D}
\end{figure}

\begin{figure}[t]
\includegraphics[width=0.95\columnwidth]{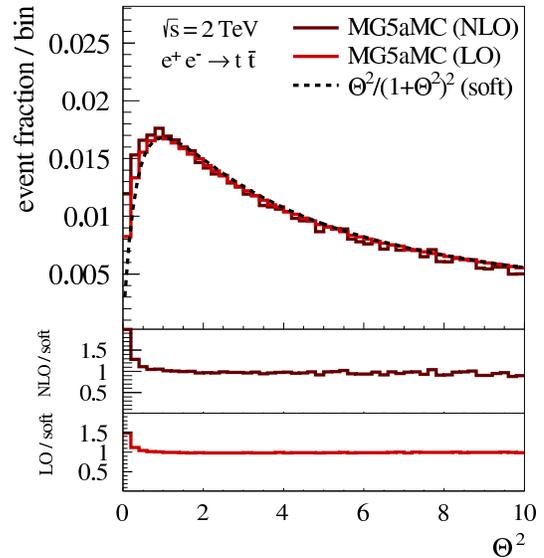}
\caption{Idealized distributions for $\Theta^2 = X^2 + Y^2$ in $e^+e^- \to t \bar{t}$ at $\sqrt{s} = 2~\TeV$, comparing LO and NLO calculations to the universal form in \Eq{eq:dc}.}
\label{fig:ee_tt_1D_ideal}
\end{figure}

\begin{figure}[t]
\subfloat[]{\label{fig:ee_tt_1D_py}
\includegraphics[width=0.95\columnwidth]{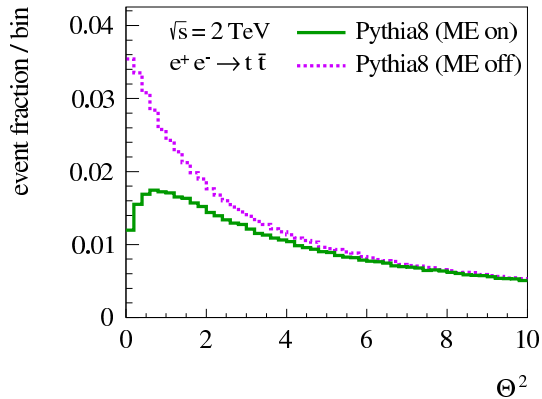}
}

\subfloat[]{\label{fig:ee_tt_1D_ps}
\includegraphics[width=0.95\columnwidth]{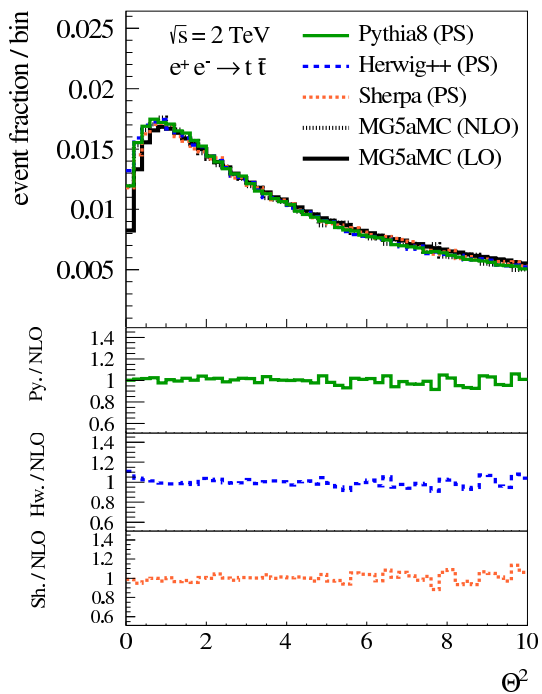}
}
\caption{The same as \Fig{fig:ee_tt_1D_ideal}, but now considering PS predictions.  (a) Turning ME corrections off and on in \pythia{}.  With ME corrections off, an erroneous collinear peak  is seen at $\Theta^2 = 0$. With ME corrections on, the dead cone is apparent for $\Theta^2 \lesssim 1$. (b) Comparing \pythia{}, \herwig{}, and \sherpa{} PS generators to LO and NLO fixed-order calculations.}
\label{fig:ee_tt}
\end{figure}

To visualize the dead cone, in \Fig{fig:ee_tt_2D} we show the full matrix element (ME) at next-to-leading order (NLO) from \aNLO{2.3.2} (MG5aMC)~\cite{Alwall:2014hca}, i.e.~$\mathcal{O}(\alpha_s^2)$ with up to two additional final-state partons (typically two gluons).  Here, we have defined
\be
\label{eq:rescaled}
\Theta \equiv \frac{\theta}{\theta_D}, \qquad X \equiv \Theta \cos \phi, \qquad Y \equiv  \Theta \sin \phi,
\ee
such that the top flight direction is at $(X,Y) = (0,0)$ and the dead cone peak is at $\Theta^2 = X^2 + Y^2 \simeq 1$.  The effective $t^* \to t g$ kinematics are determined by forcing the top quark to be stable and taking the ``gluon'' to be the vector sum of all radiated particles that are closer to the top than to the anti-top, effectively partitioning the event into top and anti-top hemispheres.\footnote{Different definitions of the ``gluon'' will change the precise shape of the dead cone.  We use jet substructure techniques in \Sec{sec:strategy}, starting from a jet cone of radius $R = 1.0$.}  This distribution is for $\sqrt{s} = 2~\TeV$ after imposing a cut on the ``gluon'' of $E_g > 50~\GeV$.

In \Fig{fig:ee_tt_1D_ideal}, we plot the idealized analytic distribution in \Eq{eq:dc} together with the corresponding distributions obtained from the exact LO and NLO fixed-order calculations for $e^+e^- \to t \bar{t} j$. This comparison shows that the dead cone radiation pattern is stable under radiative corrections and agrees well with the analytic approximation in \Eq{eq:dc}.  The deviations at LO near $\Theta^2 \simeq 0$ can be largely attributed to the $E_g > 50~\GeV$ cut, which forces us away from the strict soft limit (as motivated by the discussion around \Eq{eq:width2} below).  Though not shown here, we tested that the expected $1/z$ behavior in \Eq{eq:dc} is also seen in the fixed-order calculations.

As a next step towards a realistic modeling of the radiation from a top quark, we consider the impact of multiple gluon emissions using PS generators.  In \pythia{8.219}~\cite{Sjostrand:2006za,Sjostrand:2007gs}, the dead cone effect is implemented via ME corrections \cite{Norrbin:2000uu}.  These ME corrections can also be turned off, an option we exploit later to define a null test.\footnote{We thank Torbj\"{o}rn Sj\"{o}strand for resolving a bug in \pythia{8.215} that obscured the dead cone effect.  The proper ME corrections are applied from \pythia{8.219} on.}  In \Fig{fig:ee_tt_1D_py}, we compare the \pythia{} distributions for $\Theta^2$ with and without the dead cone effect.  One can clearly see the collinear peak at $\Theta^2 = 0$ when the corrections are off and the expected dead cone suppression for $\Theta^2 \lesssim 1$ when the corrections are on.   In this plot, the distributions have a common normalization such that the ME-corrected distribution integrates to unity.

In \Fig{fig:ee_tt_1D_ps}, we compare \pythia{8.219}~\cite{Sjostrand:2006za,Sjostrand:2007gs} to  \herwigpp{2.7.1}~\cite{Bahr:2008pv} and \sherpa{2.2.0}~\cite{Gleisberg:2008ta} as well as to fixed-order LO and NLO distributions from \aNLO{2.3.2}~\cite{Alwall:2014hca}.  In this case, we normalize the distributions to unity to emphasize any possible shape differences.  The predictions from PS generators feature the dead cone suppression for $\Theta^2 < 1$, in quite good agreement (better than 10\%) with NLO fixed-order predictions, clearly displaying the expected universal behavior.


\section{Top decay and contaminating radiation}
\label{sec:width}

We now pass from the idealized case above where the top quark has been treated as stable to the realistic case involving effects due to its decay. The top quark has a very short lifetime and decays almost exclusively to a bottom quark and a weak boson ($t \to b W$). The $W$ boson has a large ($\simeq 68\%$) branching fraction to hadronic final states, yet in order to avoid unnecessary further contamination of the dead cone, we focus on leptonic decays ($W \to \ell \nu$).  Because the $b$ quark is a colored particle, its contamination of the top dead cone due to radiation and fragmentation is unavoidable and needs to be carefully examined.

\begin{figure}[t]
\subfloat[]{\label{fig:radprod}
\includegraphics[width=0.5\columnwidth]{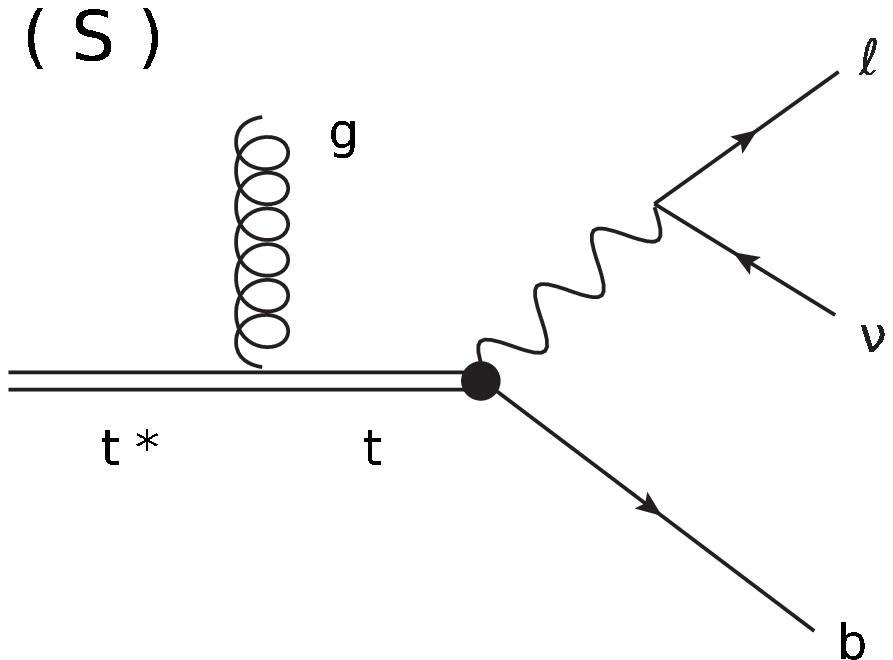}
}

\subfloat[]{\label{fig:raddec1}
\includegraphics[width=0.95\columnwidth]{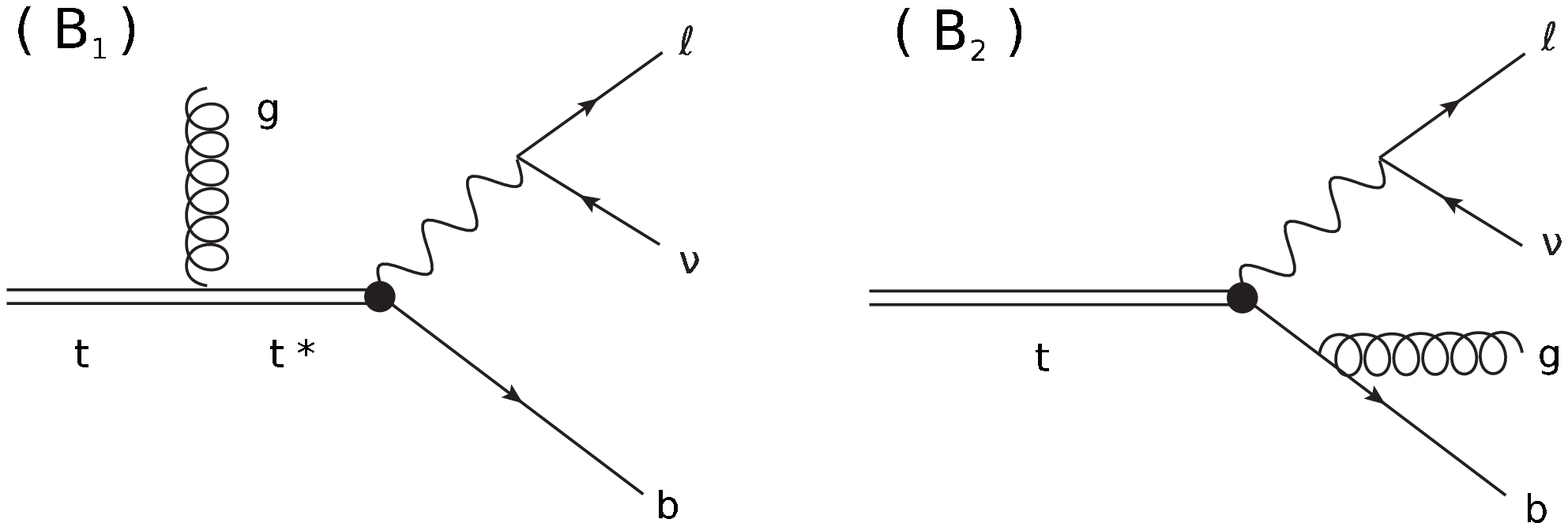}
}
\caption{Feynman diagrams for gluon radiation in (a) the signal process of top FSR $t^* \to t g$ and (b) the background process of top decay $t \to b W g$.}
\label{fig:diag}
\end{figure}

At leading order, two different gluon emission processes can be identified, as shown in \Fig{fig:diag}.
The signal process which features the dead cone is FSR top-quark radiation, corresponding to an off-shell top emitting a gluon and going on shell (see  \Fig{fig:radprod}):
\be
S: \quad t^* \to t g.
\ee
This is the process that defines the dead cone distribution in \Eq{eq:dc}.  The background process where the dead cone is absent is gluon emission during on-shell top decay (see \Fig{fig:raddec1}):
\be
B_{1,2}: \quad t \to b W g.
\ee
Even though diagrams $S$ and $B_1$ both have an off-shell top propagator, $B_1$ does not contribute to the dead cone effect.  This is easiest to see in limit that the top quark is exactly stable.  Specifically, one can boost to the on-shell top rest frame, where it is then apparent that the emitted gluon in diagram $B_1$ is uncorrelated with the initial top momentum direction.  Because the top is unstable, though, there is interference between the signal and background processes proportional to the top-quark width $\Gamma_{t} \simeq 1.4~\GeV$.  This interference becomes relevant when
\begin{equation}
2 \, p_t\cdot p_g \sim m_{t} \Gamma_{t},
\label{eq:width}
\end{equation}
where $p_t$ and $p_g$ are the top and gluon 4-momenta.  Indeed, observables have been proposed to exploit this interference regime and potentially measure the top-quark width~\cite{Dokshitzer:1992nh,Orr:1993kd,Khoze:1993ij}.

Here, our goal is to isolate the $S$ process, so we want to avoid interference effects.   In addition, if the gluon energy is too small, then there is no practical way to distinguish an on-shell top from an off-shell top, allowing the $B_{1,2}$ diagrams to ``bleed through'' into the $S$ diagram signal region. To estimate when interference effects can be neglected, we use the relation $2 \, p_t\cdot p_g  \sim E_t E_g \theta_D^2$ for sufficiently small angles in the lab frame, leading to the requirement
\begin{equation}
z \equiv  \frac{E_g}{E_t} \gg \frac{\Gamma_{t}}{m_t},
\label{eq:width2}
\end{equation}
which for $\Gamma_{t}/{m_t} \simeq 0.01$ implies $z \gtrsim \mathcal{O}(0.1)$.   In \App{app:interference}, we explicitly check that \Eq{eq:width2} with $z > 0.05$ is sufficient to suppress the combined interference and bleed-through effects.  We consider further kinematic selections to suppress the decay processes in the next section.

We conclude this section by commenting on additional sources of background that are present in $pp$ collisions coming from ISR and UE.  Radiation associated with the initial partons involved in the scattering as well as with soft QCD effects from the proton remnants can ``accidentally'' end up in the vicinity of the reconstructed top quark.  Because top quarks are dominantly produced via gluon fusion at the LHC, and because the degree of ISR is controlled by the gluon color factor $C_A = 3$, ISR turns out to be a rather important source of dead cone contamination.  In addition, UE contributes to an overall pedestal in the $\Theta^2$ distribution which also fills in the dead cone region.  This motivates the use of jet grooming techniques to mitigate the impact of ISR/UE contamination.


\section{Exposing the dead cone at the LHC}
\label{sec:strategy}

We now present an analysis strategy to observe the dead cone effect at the LHC.  Our starting point is an event sample of boosted top quark pairs, with one top quark decaying hadronically and the other one leptonically. The boosted leptonic top  (BLT) is where we propose to measure the dead cone effect.

The reason for considering large Lorentz boosts is that top FSR is roughly proportional to $\alpha_s \log (E_t/m_t)$.  By going to larger values of $E_t$, the overall level of FSR is enhanced, making the dead cone suppression more distinct.  We consider top quarks with transverse momenta of $p_T \gtrsim 500~\GeV$ for which the expected dead cone angle is $\theta_D  \simeq 0.3$, safely larger than the angular resolution of the LHC detectors.

The reason for considering single-lepton top pairs is threefold.   First, identifying a boosted hadronic top ensures high signal purity when using jet substructure tagging techniques \cite{CMS:2014fya}.  Second, the single-lepton selection ensures that the primary source of missing energy comes from the single neutrino in the event, allowing an accurate reconstruction of the BLT direction.  Third, performing the measurement on the BLT avoids hadronic dead cone contamination from the decaying $W$ boson.  As discussed in \Sec{sec:width}, residual contamination of the top dead cone comes mainly from $b$-quark FSR as well as from ISR and UE.

\subsection{Event selection}

Our baseline event selection involved two large radius ($R = 1.0$) ``fat" jets  at central rapidities ($|\eta^j| < 2.5$) and high transverse momenta ($p^j_T > 300~\GeV$).  Jets are reconstructed using the anti-$k_T$ jet algorithm~\cite{Cacciari:2008gp} from \fastjet{3.1.3}~\cite{Cacciari:2011ma}.  Exactly one of the two fat jets is required to satisfy a boosted hadronic top tag ($p^t_T > 500~\GeV$), and we use \Ref{CMS:2014fya} to estimate the tagging performance.  The other jet is promoted to a BLT candidate if it contains at least one high-$p_T$ lepton ($p^\ell_T > 50~\GeV$).  To avoid misreconstructing the BLT kinematics due to collinear photon FSR, we define the effective lepton four-vector to include all photons within $\Delta R_{\gamma\ell} < 0.1$.  We further require large missing transverse momentum ($p_T^\text{miss} > 50~\GeV$).

The above selection is designed to obtain high purity of BLT signal events.  To further increase the signal yield, one could lower the top $p_T$ threshold and possibly widen the jet radius, though we found that this did not improve the statistical significance of the dead cone effect.  Additional potential backgrounds are discussed in \Sec{sec:backgrounds}, though they are expected to only appear at the few-percent level.

\subsection{Object reconstruction}

From the BLT constituents, we next need to define the $b$ quark, FSR ``gluon" ($g$), lepton ($\ell$), and neutrino ($\nu$) candidates.  In addition, we want to mitigate the impact of  contamination from ISR/UE, as well as from pileup at higher luminosities.

To identify the $b$ and $g$ candidates and suppress  contamination, we exploit recent advances in jet substructure.   After removing the high $p_T$ lepton (and its collinear photon FSR), the BLT constituents are reclustered with the Cambridge-Aachen (C/A) algorithm~\cite{Dokshitzer:1997in}, which reorganizes the BLT constituents into an angular-ordered tree.  We then apply the soft drop algorithm \cite{Larkoski:2014wba}, which aims to remove soft contamination from the fat jet and isolate two subjets within the BLT.  Soft drop works by recursively declustering the C/A tree, removing the softer branch until
\be
\frac{\min[p_{T1}, p_{T2}]}{p_{T1} + p_{T2}} > z_\text{cut} \left(  \frac{R_{12}}{R} \right)^\beta,
\ee
where $p_{Ti}$ are the transverse momenta of the subjets, $R_{12}$ is their rapidity-azimuth distance, and $R$ is the initial jet radius.  For this study, we use the parameters
\be
\beta = 0, \qquad z_\text{cut} = 0.05.
\ee
By choosing $\beta = 0$, soft drop behaves similarly to the modified mass drop tagger with $\mu = 1$ \cite{Dasgupta:2013ihk}.  In order to increase the signal acceptance, we have selected the $z_\text{cut}$ value to be a bit looser than the 0.1 value used in \Refs{Larkoski:2014wba,Larkoski:2014bia,Larkoski:2015lea}.

If soft drop finds no substructure within the hadronic component of the BLT jet, the event is discarded.  If soft drop instead finds evidence for a 2-prong substructure, then the 4-momenta of two subjet components are returned. Out of the two subjets found by soft drop, exactly one subjet is required satisfy a $b$ tag to become our $b$ quark candidate.  The remaining subjet is our ``gluon" candidate, which ideally would come from the top FSR signal.  We further impose $p_T^b > 50~\GeV$ and $p_T^g > 25~\GeV$ to avoid pathological configurations.

To identify the neutrino candidate, we use the $W$ mass constraint on the $\ell \nu$ system to solve for the missing neutrino longitudinal momentum component.  Strictly speaking, the $W$ mass constraint yields two solutions for the neutrino longitudinal momentum, so we choose the one that gives the smallest value of $\min\{|m_t - m_{b\ell\nu}|,|m_t - m_{b\ell\nu g}|\}$.  By considering both the $b\ell\nu$ and $b\ell\nu g$ systems, we avoid sculpting an artificial dead cone region.  After reconstructing the neutrino direction, we now have a BLT candidate with well-defined $b$, $\ell$, $\nu$, and $g$ constituents.

Finally, we impose a cut on the ``gluon'' momentum relative to the reconstructed BLT momentum in order to satisfy the interference and bleed-through requirement from \Eq{eq:width2}:
\be
\label{eq:egoveret}
\frac{p_T^g}{p_T^t} > 0.05.
\ee
Note that this requirement is stricter than the $z_\text{cut}$ requirement of the soft drop algorithm, which only constrains the ``gluon'' momentum relative to the momentum of the top minus $W$ system.  In practice, \Eq{eq:egoveret} is often satisfied already by the $p_T^g > 25~\GeV$ requirement, since the typical top $p_T$ for this selection is $500~\GeV$.

\subsection{Signal isolation}
\label{sec:sigiso}

\begin{figure}
\includegraphics[width=0.95\columnwidth]{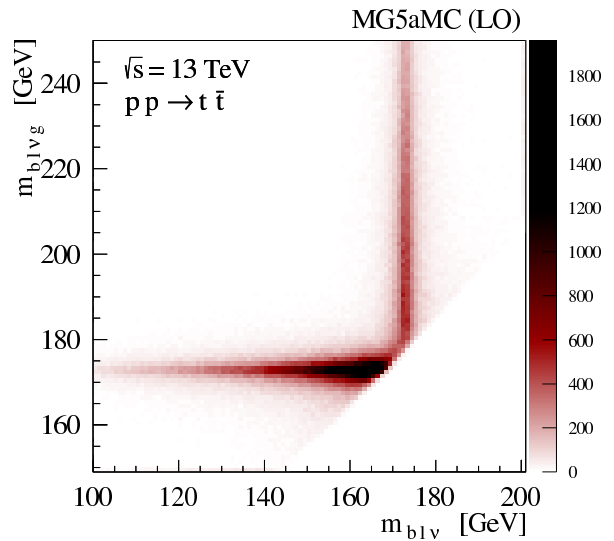}
\caption{Invariant mass of $b\ell\nu$ versus $b\ell\nu g$ after the soft drop procedure, showing the regions that are signal enriched (vertical bar) and background enriched (horizontal bar) at LO.}
\label{fig:pp_mass2D_fo}
\end{figure}

\begin{figure*}[t]
\subfloat[]{\label{fig:pp_tt_sg_2D}
\includegraphics[width=0.95\columnwidth]{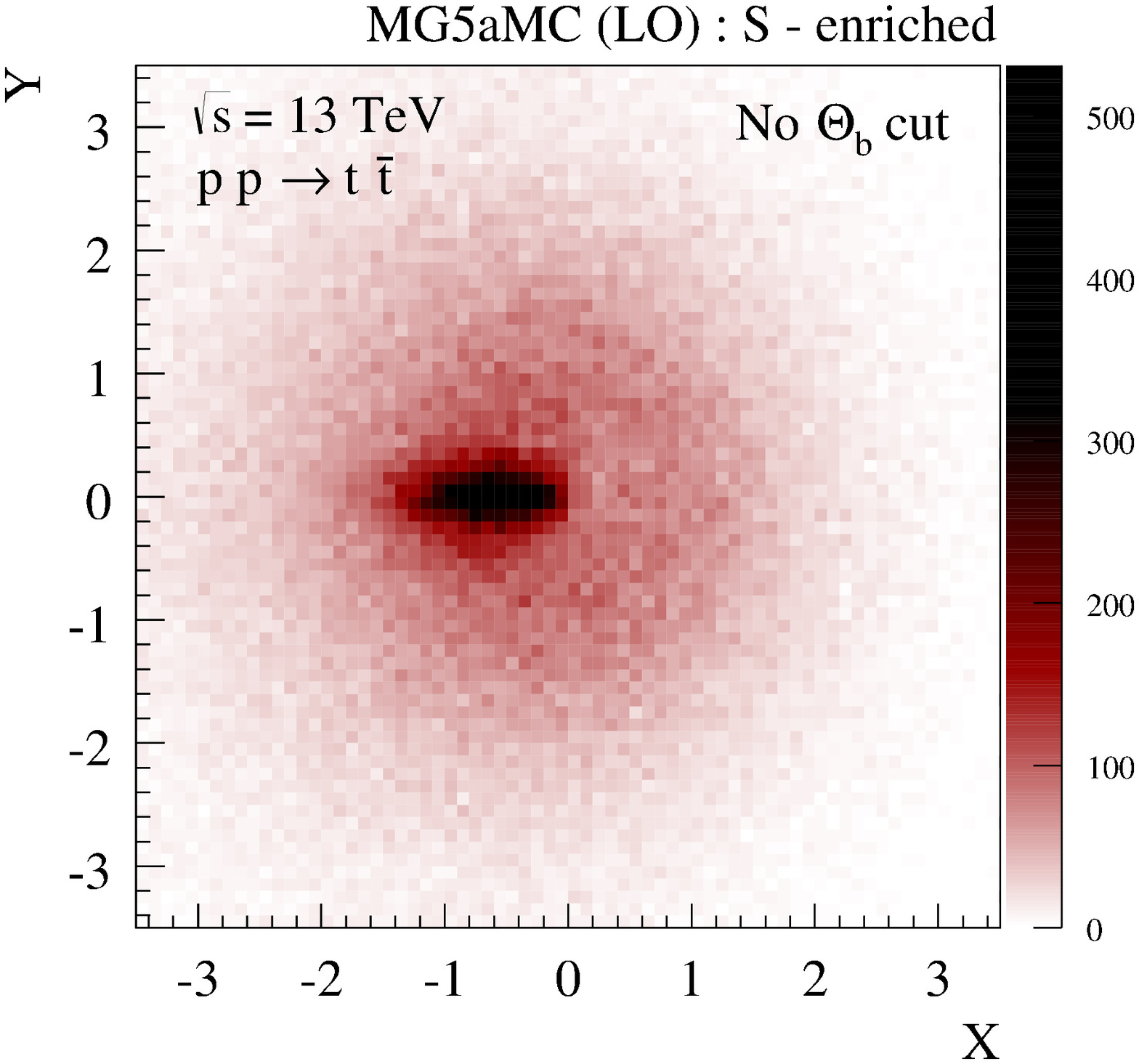}
}
$\quad$
\subfloat[]{\label{fig:pp_tt_bk_2D}
\includegraphics[width=0.95\columnwidth]{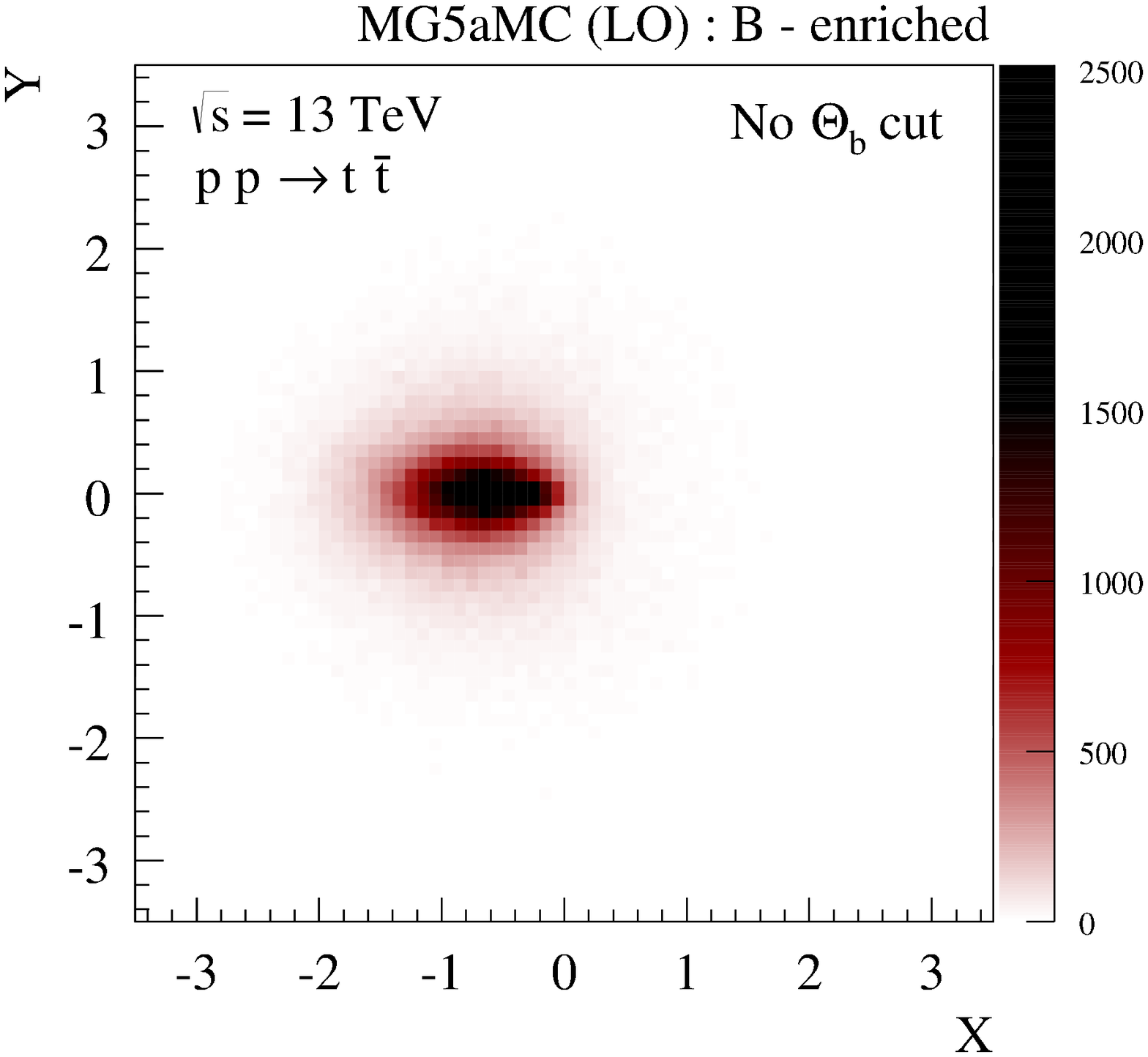}
}
\caption{Realistic angular distribution for gluon radiation in 13 TeV LHC collisions at LO.   The event is rotated such that the reconstructed top flight direction is at $(X,Y) = (0,0)$ and the $b$-jet points along the negative $X$-axis.  (a) The $S$-enriched region ($m_{b\ell\nu} \in [170,200]~\GeV$), showing a disk of top FSR within $\Theta^2 \lesssim 1$ along with contamination from $b$-quark FSR at $X < 0$.  (b) The $B$-enriched region ($m_{b\ell\nu} < 160~\GeV$) where no disk-like top FSR pattern is expected (or observed).  See \Fig{fig:pp_tt_2D_Th10} for further cuts to isolate the top dead cone suppression at $\Theta = 0$.}
\label{fig:pp_tt_2D}
\end{figure*}

With the BLT kinematics in hand, we now take advantage of the differing kinematics in top FSR and top decay.  When the gluon is radiated in top FSR then $m_{b\ell\nu} \approx m_t$ and $m_{b\ell\nu g} > m_t$, whereas when the gluon is radiated in top decay then $m_{b\ell\nu} < m_t$ and $m_{b\ell\nu g} \approx m_t$.  These two regions can be seen clearly in \Fig{fig:pp_mass2D_fo}, from a LO calculation where all of the above selection criteria are applied.\footnote{\label{footnote:thadsubtle}For reasons of computational efficiency, this LO calculation is for $pp \to t_{\rm had} b \ell \nu g$, where $t_{\rm had}$ refers to a hadronic top quark that is treated as stable.  While strictly speaking not gauge invariant when the leptonic top is off shell, this amplitudes provides an excellent approximation to the full one, with a negligible uncertainty in our analysis.}

We therefore have a signal-enriched phase space region where the dead cone effect should be enhanced and a background-enriched control region where no dead cone is expected:
\begin{align}
\mbox{$S$-enriched:} &\quad  m_{b\ell\nu} \in [170,200]~\GeV,\\
\mbox{$B$-enriched:} &\quad  m_{b\ell\nu} < 160~\GeV.
\end{align}
While one might try to cut on $m_{b\ell\nu g}$ to further enhance top FSR and suppress top decay in the $S$-enriched region, we find that this sculpts an artificial dead cone since it preferentially selects events with wide-angle gluons.

At this point, it is convenient to rotate the event such that the momentum of the reconstructed BLT (including the radiated gluon) points in the \emph{z} direction, and the $b$-subjet candidate has a vanishing \emph{y} component and negative $x$ component.  This allows us to isolate (and better visualize) background-like configurations where the gluon candidate is likely to come from $b$-quark FSR ($x<0$, dominated by diagram $B_2$) from signal-like configurations where the gluon candidates is likely to come from top-quark FSR ($x>0$, diagram $S$).  As in \Eq{eq:rescaled}, we rescale the gluon kinematics to $X$ and $Y$ coordinates; this ensures that the expected dead cone boundary is at $\Theta^2 = 1$ regardless of the reconstructed top momentum.\footnote{To better match the discussion in \Sec{sec:leptoncase}, we define the $\Theta$ coordinate in terms of lab-frame energies and angles, instead of the more familiar $p_T$ and $\Delta R$.  The difference is negligible for narrow jets and only leads to a small distortion for the $R =1.0$ jet radius used here.}

The resulting gluon radiation pattern is shown in \Fig{fig:pp_tt_2D}.  In both the $S$- and $B$-enriched samples, there is a prominent peak near $(X,Y) = (-1,0)$, corresponding roughly to the $b$ quark location.  This peak is expected, since even with the $S$-enriched selection, there is still residual contamination from $b$-quark FSR.  For the $S$-enriched sample there is a faint disk of radiation within $\Theta^2 \lesssim 1$, though scant evidence for dead cone depletion near the $\Theta = 0$ origin.  This disk corresponds to the desired top FSR signal seen in the idealized distribution from \Fig{fig:ee_tt_2D}.  No such disk-like feature is observed in the $B$-enriched sample, giving us confidence that the $S$-enriched selection has properly isolated the top FSR of interest.

\begin{figure*}[t]
\subfloat[]{\label{fig:pp_tt_sg_2D_Th10}
\includegraphics[width=0.95\columnwidth]{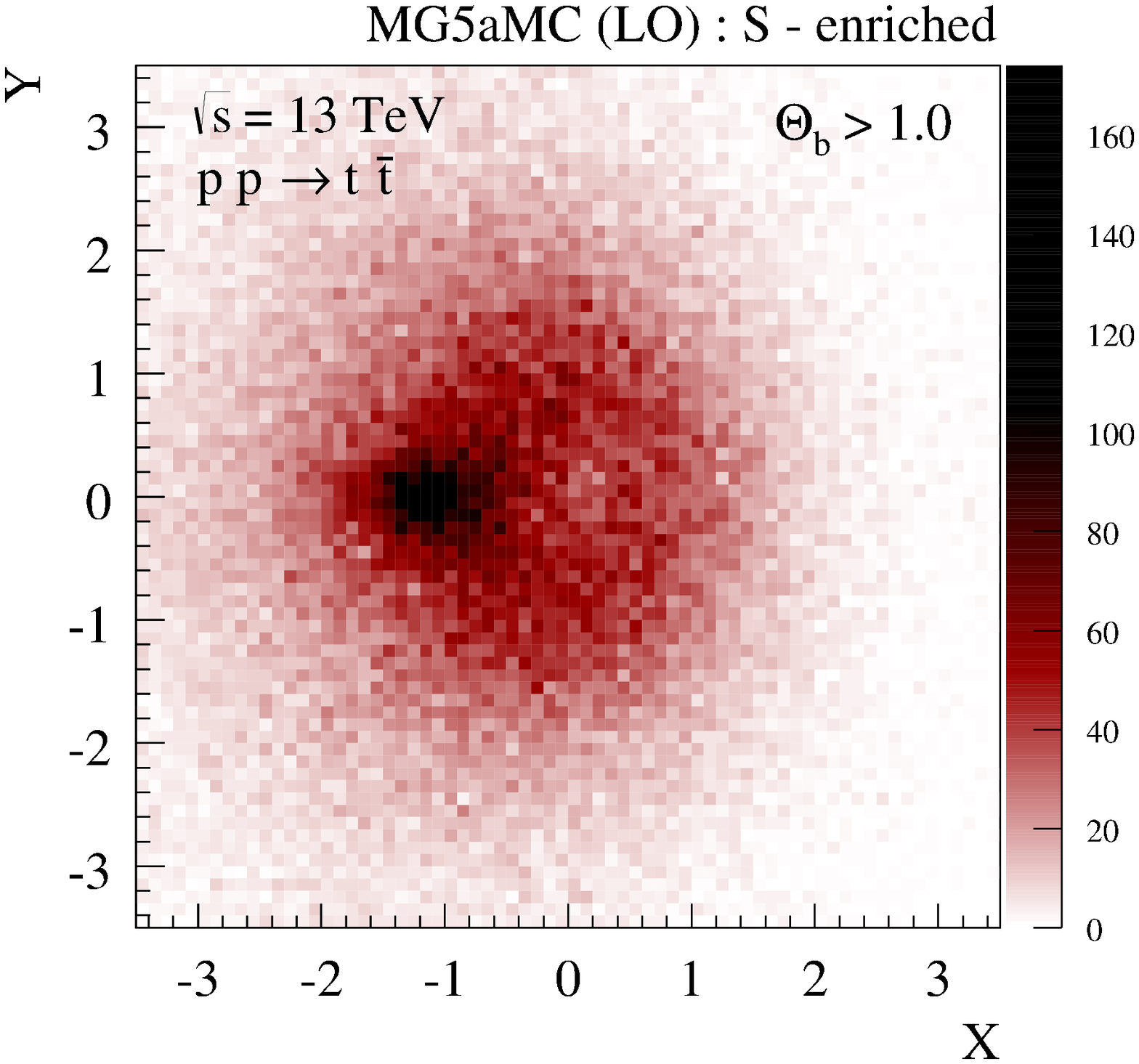}
}
$\quad$
\subfloat[]{\label{fig:pp_tt_bk_2D_Th10}
\includegraphics[width=0.95\columnwidth]{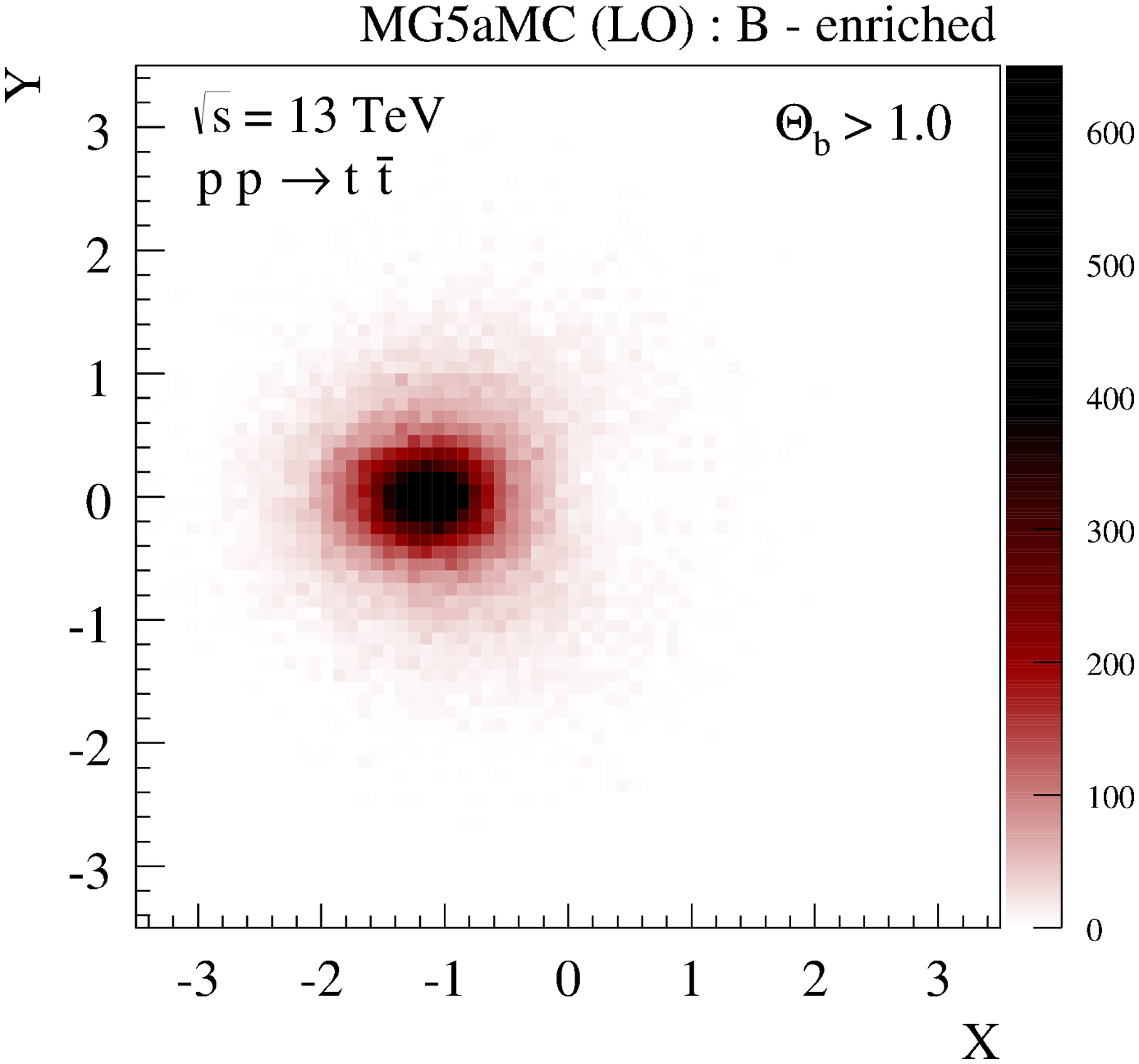}
}
\caption{The same as \Fig{fig:pp_tt_2D}, but applying a cut on the $b$ quark candidate of $\Theta_b > 1.0$.  The dead cone effect at $\Theta = 0$ is noticeably enhanced in the $S$-enriched region without sculpting a feature in the $B$-enriched region.}
\label{fig:pp_tt_2D_Th10}
\end{figure*}

In \App{app:thetab}, we show distributions for the observable
\be
\label{eq:rescaledX}
\Theta_S^2 \equiv \mathrm{sign}(X) \, \Theta^2,
\ee
such that $\Theta_S^2 > 0$ isolates the phase space region away from the $b$ quark direction.  However, with no further cuts, the realistic dead cone structure in \Fig{fig:pp_tt_sg_2D} is rather muted compared to the idealized dead cone structure in \Fig{fig:ee_tt_2D}.  One might therefore wonder if there are additional kinematic handles to enhance the dead cone effect and observe a suppression near $\Theta = 0$.

\subsection{Further optimization}
\label{sec:thetab}

\begin{figure*}[t]
\subfloat[]{\label{fig:pp_tt_sg_1D_Th10}
\includegraphics[width=0.95\columnwidth]{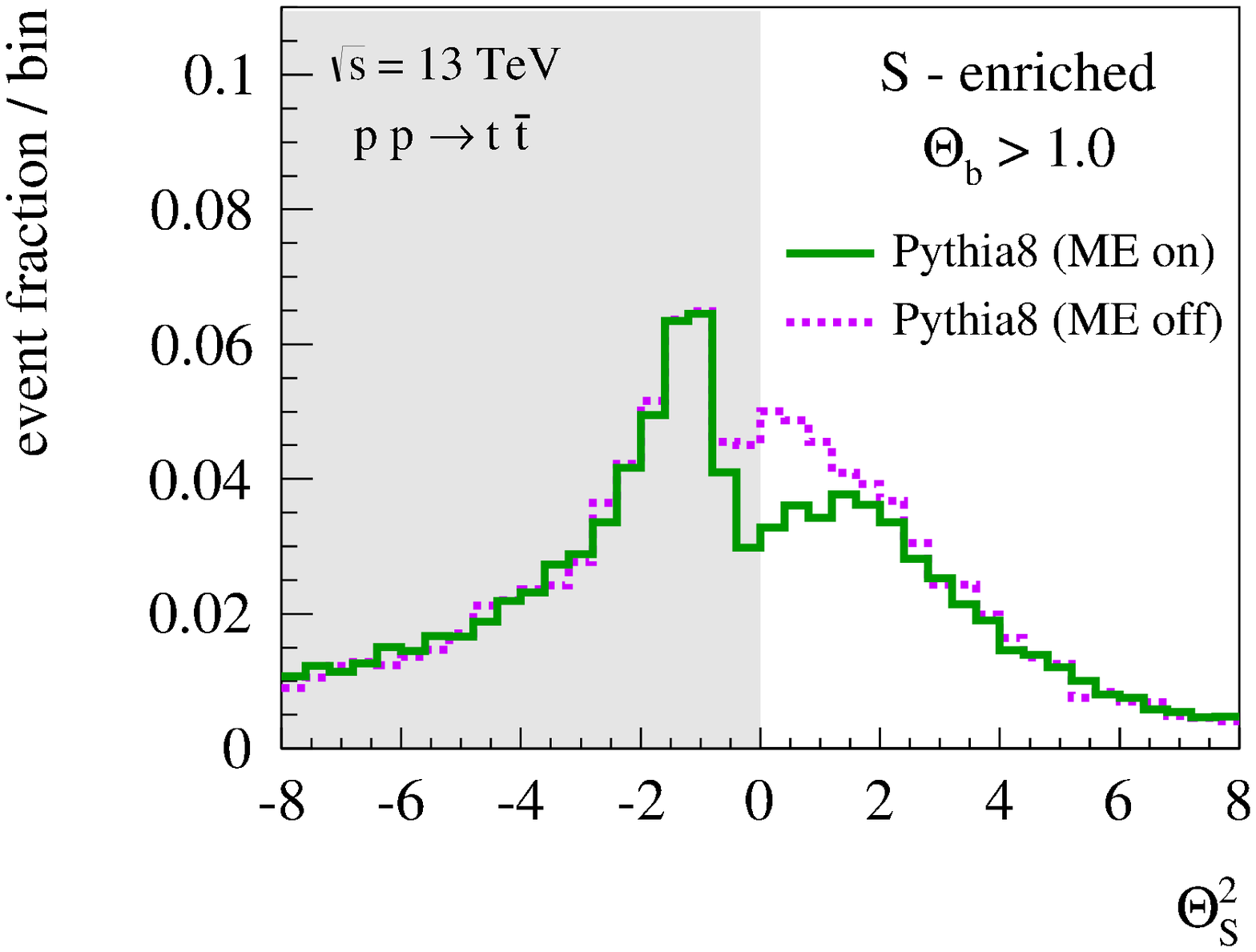}
}
$\quad$
\subfloat[]{\label{fig:pp_tt_bk_1D_Th10}
\includegraphics[width=0.95\columnwidth]{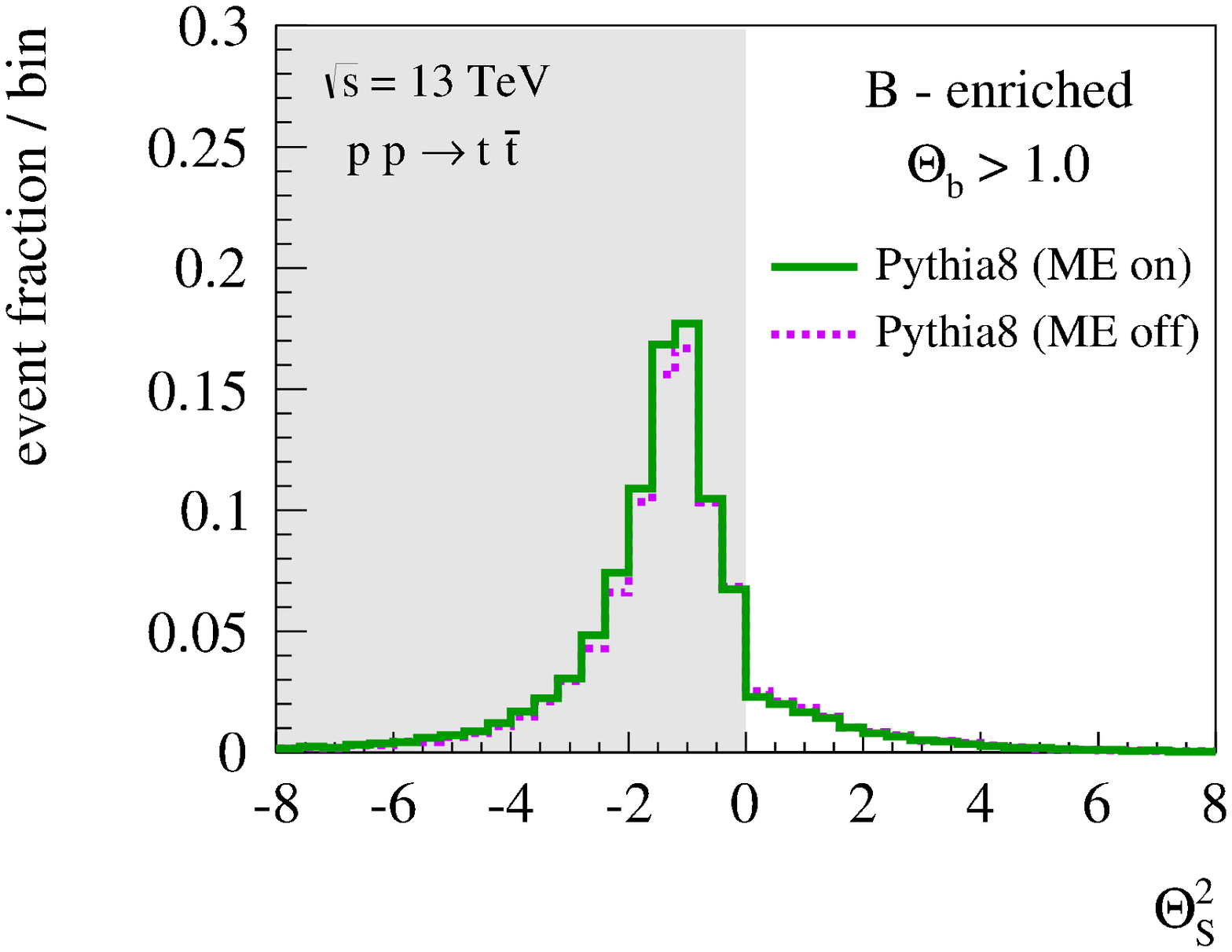}
}

\subfloat[]{\label{fig:pp_tt_sg_1D_ps_Th10}
\includegraphics[width=0.95\columnwidth]{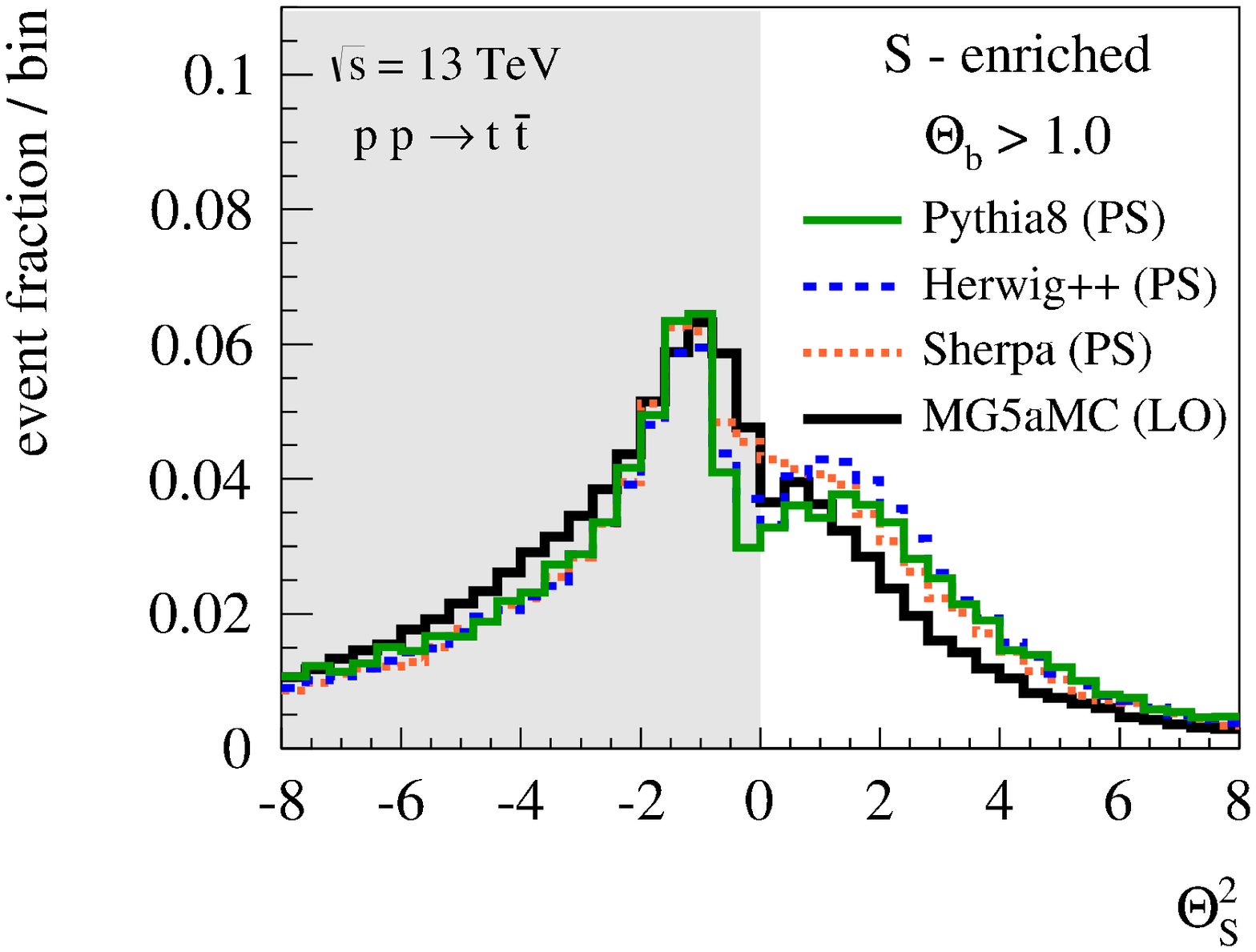}
}
$\quad$
\subfloat[]{\label{fig:pp_tt_bk_1D_ps_Th10}
\includegraphics[width=0.95\columnwidth]{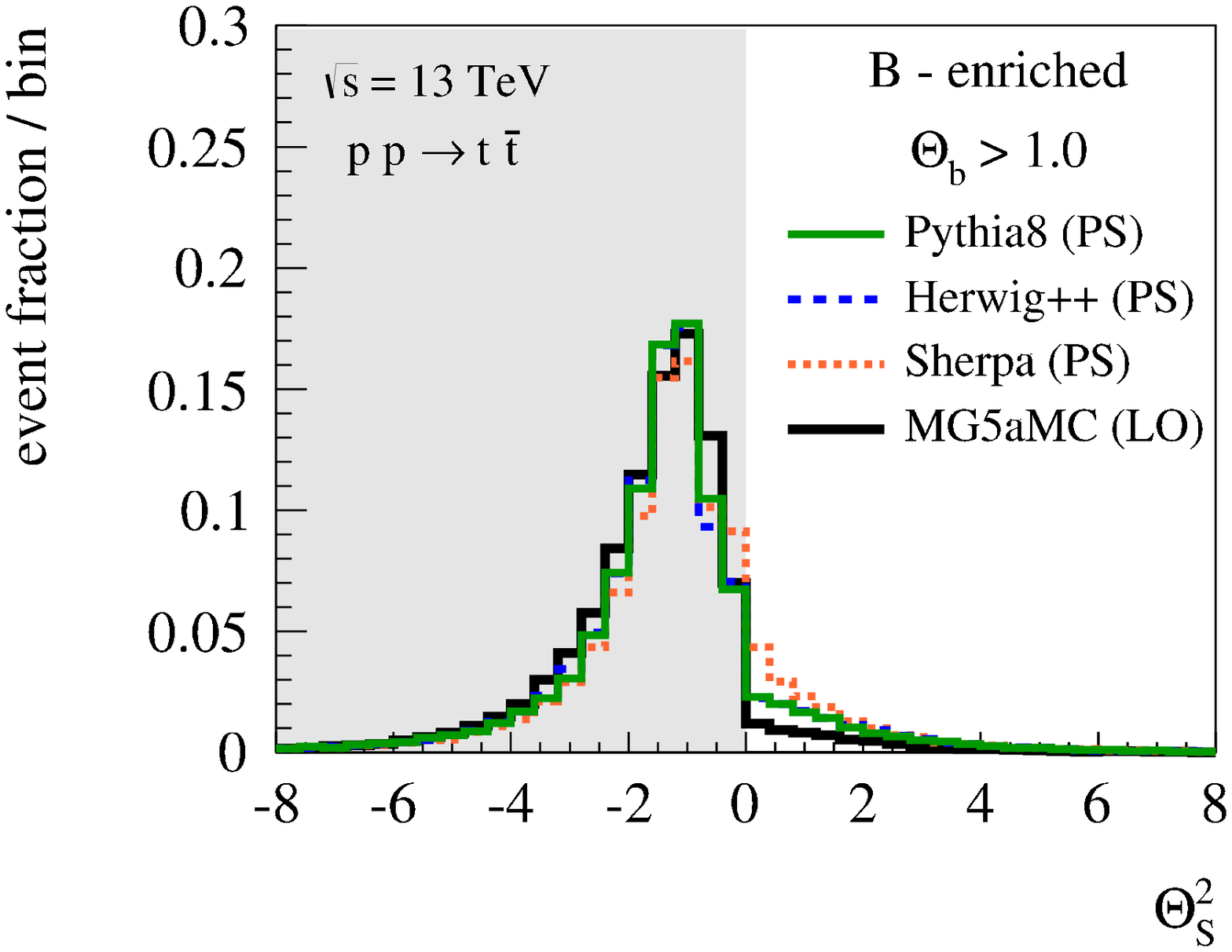}
}
\caption{Realistic distributions for $\Theta_S^2 = \mathrm{sign}(X) (X^2 + Y^2)$ in 13 TeV LHC collisions, for the (left column) $S$-enriched and (right column) $B$-enriched samples.  Here, a cut of $\Theta_b > 1.0$ has been applied; see \Fig{fig:pp_tt} for distributions without this cut.  (Top row)  Turning ME corrections off and on in \pythia{}.  Restricting one's attention to $\Theta_S^2 \in [0,1]$ for the $S$-enriched sample, one sees qualitatively the same dead cone physics as in \Fig{fig:ee_tt_1D_py}.  ME corrections have a negligible impact for the $B$-enriched sample.   (Bottom row) Comparing the PS generators to LO fixed-order calculations.  While each generator shows some evidence for a dead cone suppression, the quantitative behavior is noticeably different.}
\label{fig:pp_tt_Th10}
\end{figure*}

The most obvious source of dead cone contamination is $b$-quark FSR.  By two-body kinematics, the typical opening angle between the $b$-quark direction and the initial top-quark direction is typically \emph{the same} as the dead cone angle, $\theta_{tb} \approx \theta_D$.  Moreover, the total gluon FSR from the $b$ quark is expected to be larger than from the top quark.\footnote{Bottom FSR is proportional to $\alpha_s \log (E_b^*/m_b)$ where $E_b^*$ is the bottom energy in the top rest frame, while top FSR is proportional to $\alpha_s \log (E^*_t/m_t)$ where $E^*_t$ is the top energy in the $t \bar{t}$ rest frame, so the former dominates at the moderate top boosts considered here.}  While imposing a $\Theta_S^2 > 0$ restriction could help isolate the phase space region away from the $b$ quark, a more aggressive way to ``clean up'' the dead cone region is to force the $b$ candidate to have a large value of $\Theta_b$.

In \Fig{fig:pp_tt_2D_Th10}, we show the impact of a $\Theta_b > 1.0$ restriction.  Such a cut (which could be optimized in a full analysis) ensures that $b$-quark FSR is typically away from the dead cone region $\Theta^2 \lesssim 1$.  Because this does not impose any criteria on the gluon subjet, this selection does not sculpt a dead cone, though it does preferentially select top decays where the $b$ quark flies perpendicular to the top boost direction in the top rest frame.

With this $\Theta_b > 1.0$ cut in place, we project down to the $\Theta_S^2$ observable from \Eq{eq:rescaledX} in \Fig{fig:pp_tt_Th10}.  Focusing on $\Theta_S^2 > 0$ for the $S$-enriched sample in \Fig{fig:pp_tt_sg_1D_Th10}, the dead cone structure is quite visible.  When turning the ME corrections off in \pythia, one sees a rise towards $\Theta_S^2 = 0$, corresponding to FSR emitted collinear to the initial top direction.  With the ME corrections on, the dead cone region at $\Theta_S^2 \lesssim 1$ appears as expected.  No such features are seen in the $B$-enriched sample in \Fig{fig:pp_tt_bk_1D_Th10}, suggesting that $\Theta_S^2$ is a useful test for the dead cone effect at the LHC, especially after a cut on $\Theta_b$.  Comparing different predictions in \Fig{fig:pp_tt_sg_1D_ps_Th10}, we see that each generator predicts some degree of dead cone suppression for $\Theta_S^2 \lesssim 1$, though the precise size and shape differs noticeably, motivating future higher-order calculations of the dead cone effect in $pp$ collisions.  Even without new calculations, these generator differences indicate that a direct measurement of the dead cone effect at the LHC would help test Monte Carlo treatments of gluon radiation from massive quarks.

One particular challenge in $pp$ collisions that is absent from $e^+e^-$ collisions is ISR and UE.  This background has no preferred orientation with respect to the top flight direction and simply leads to uniform contamination of the dead cone region, which is only partially mitigated by soft drop.  In \App{app:eereal}, we show $e^+e^- \to t \bar{t}$ collisions with the identical event selection as used in the $pp$ case, where the dead cone effect is more readily visible.  Though not shown here, we also tested our analysis strategy on just the $q \bar{q} \to t \bar{t}$ subprocess where ISR contamination is somewhat suppressed, yielding results that are intermediate between the $e^+e^-$ and full $pp$ distributions.

\section{Estimated LHC sensitivity}
\label{sec:reach}

Because the top dead cone in \Fig{fig:pp_tt_sg_1D_Th10} is still rather subtle, large data samples will be necessary to find conclusive evidence for this effect.  For an integrated luminosity $\mathcal{L}$ and signal efficiency $\epsilon_{\rm total}$, the expected number of events $\mathcal{N}$ contributing to the $S$-enriched sample can be expressed as
\begin{align}
\mathcal{N} &= \mathcal{L} \, K \, \sigma^{\rm LO} (p p \to t \bar{t}, p^{t,\bar{t}}_T > 500~\GeV) \nonumber \\
 & \quad ~ \times \mathcal{B}(t\bar{t} \rightarrow t_{\rm had}t_{\rm lep}) \, \epsilon_{\rm total}.
\label{eq:nevents}
\end{align}
Here, $\sigma (p p \to t \bar{t}, p^{t,\bar{t}}_T > 500~\GeV) = 1.4~\pb$ is the boosted top cross section at LO, $K = 1.65$ is the ratio of the inclusive 13$~\TeV$ $p p \rightarrow t\bar{t}$ cross sections at NNLO~\cite{Czakon:2013goa} compared to LO, and $\mathcal{B}(t\bar{t} \rightarrow t_{\rm had} t_{\rm lep}) = 0.30$ is the fraction of top-quark pairs featuring a single-lepton final state.

The total signal efficiency can be expressed as
\begin{equation}
\epsilon_{\rm total} \equiv \epsilon_{\rm fid} \, \epsilon_{\rm top} \, \epsilon_{\rm SD} \, \epsilon_{ b} \, \epsilon_{\Theta_b} \, \epsilon_{S}.
\end{equation}
Using \pythia{}, we estimate the efficiency of the fiducial cuts ($p^j_T > 300~\GeV$, $p^t_T > 500~\GeV$, $|\eta^{j,t,\ell}| < 2.5$, $p_T^\text{miss} > 50~\GeV$, and $p^{\ell}_T > 50~\GeV$) as $\epsilon_{\rm fid} = 45\%$.  We assume hadronic-top-tagging~\cite{CMS:2014fya} and $b$-tagging efficiencies~\cite{Chatrchyan:2012jua} of $\epsilon_{\rm top} = \epsilon_{b} = 50\%$.\footnote{The corresponding mistag rates from these CMS studies are $\epsilon_{\rm top}^{\rm mis} = 5\%$ and $\epsilon_{b}^{\rm mis} = 1\%$.}  From the same \pythia{} sample, we estimate that the soft drop tagging efficiency is $\epsilon_{\rm SD} = 55\%$, which includes the $p^{b}_T > 50~\GeV$, $p^g_T > 25~\GeV$, and $p_T^g/p_T^t > 0.05$ requirements. The efficiency for the cut $\Theta_b > 1.0$ is $\epsilon_{\rm \Theta_b} = 30\%$, and the $S$-enriched fraction is $\epsilon_{S} = 30\%$.  This gives an overall signal efficiency of
\be
\epsilon_{\rm total} \simeq 0.55\%
\ee
before placing any restrictions on $\Theta_S^2$.

For the expected integrated luminosity of $\mathcal{L} = 300~\text{fb}^{-1}$ to be collected in Run II and III of the LHC by the ATLAS~\cite{Aad:2009wy} and CMS~\cite{Bayatian:2006zz} experiments, we find $\mathcal{N} \approx \text{1150}$ top dead cone candidates.  In the crucial phase space region $\Theta_S^2 \in [0.0,1.0]$, the estimated yield is
\begin{align}
\mathcal{N}^{[0.0,1.0]}_{\rm on} &= 85,\\
\mathcal{N}^{[0.0,1.0]}_{\rm off} &= 125,
\end{align}
from \pythia{} with and without ME corrections respectively.  The difference between these yields is statistically significant at $\approx 4\sigma$, and the dead cone should be definitively testable with $300~\text{fb}^{-1}$ of LHC data.\footnote{Taking a wider interval of $\Theta_S^2 \in [-0.5,1.5]$, the estimated yield is $\mathcal{N}^{[-0.5,1.5]}_{\rm on} = 175$ and $\mathcal{N}^{[-0.5,1.5]}_{\rm off} = 235$, which also differs at a significance of $\approx 4\sigma$.  Note however, that the dip in \pythia{} for $\Theta_S^2 \in [-0.5,0.0]$ is not seen in the other generators.}   A precision differential measurement of the $\Theta_S^2$ spectrum would be possible at higher luminosities, i.e.~HL-LHC.

\section{Background Considerations}
\label{sec:backgrounds}

Our analysis thus far has assumed that the dominant background to the dead cone effect is simply $b$-quark FSR from true semi-leptonic top pair events.  That said, secondary backgrounds can arise from single top, $W$ plus jets, and all-hadronic top pairs.  These would have to be carefully considered in a full LHC analysis, especially after considering pileup and detector effects, though we estimate here that such backgrounds are negligible.

Single top production actually provides an additional source of signal events if the single top decays leptonically and the recoiling system is mistagged as a boosted hadronic top.  If the single top decays hadronically, though, there is a potential source of background events if the recoiling system consists of a (mis)tagged $b$ jet and a $W$ boson.  For example, this can occur in $t$-channel single-top events where the recoiling jet collinearly radiates a leptonic $W$, or in associated production of single top with a leptonic $W$ boson where there is an additional jet from ISR.  Given that the single top cross section is already much smaller than the $t\bar{t}$ cross section, though, such backgrounds yield a sub-percent contribution to the total event rate.

For $W$ plus jets, the LO fiducial cross section for $W+b/c+\text{jet}$ with $p_T^{j} > 500~\GeV$ and $|\eta^{j,b}| < 2.5$ is estimated to be $\sigma_{W b j} \, \mathcal{B} (W \rightarrow \ell \nu) =$ 80 fb with \aNLO{} before any event selection.  Note that this cross section incorporates the dominant contribution from dijets, which arises when a boosted dijet system undergoes electroweak FSR, leading to the $W$ plus jets final state already considered.  Requiring the light jet to pass a top tag reduces the contribution of this background to the sub-percent level.

All-hadronic top pairs are a potential background if the $b$ quark decays semi-leptonically and one of the $W$ decay products fakes the top radiation.  This background can be estimated from simulation, and with the cuts of $p^{b,\ell}_T > 50~\GeV$ and $p_T^\text{miss} > 50~\GeV$, we estimate that it should contribute at most at the few-percent level.  Despite being small, all-hadronic top pairs are likely to be the most important secondary background to consider in a full analysis.

Finally, we note that our $p_T^g > 25~\GeV$ and $p_T^g/p_T^t > 0.05$ cuts might be too loose given the possibility of pileup jets at high luminosity.\footnote{Recall, though, that the value of $p_T^g$ is inferred after performing the soft drop procedure, which would certainly help to mitigate the effect of pileup jets.}  The reason for our cut choice is that the dead cone effect is most robust in the soft gluon limit, up until the point where the interference and bleed-through effect from \Eq{eq:width2} becomes relevant.  We checked that the qualitative features of our analysis still persist with a soft drop parameter of $z_{\rm cut} = 0.1$ and harder cuts of $p_T^g > 50~\GeV$ and $p_T^g/p_T^t > 0.1$, though the statistical significance of the signal  with $300~\text{fb}^{-1}$ is degraded down to the $2\sigma$--$3\sigma$ level.  If a tighter $p_T^g$ cut is needed, then one would likely want to revisit the $p^{b,\ell}_T$ and $p_T^\text{miss}$ requirements as well, especially if there are alternative methods available to suppress the all-hadronic top background.

\section{Conclusions}
\label{sec:conclude}

With the excellent performance of the ATLAS and CMS detectors, the high luminosities foreseen at the LHC, together with new analysis techniques based on jet substructure, there is an opportunity to study subtle physics effects involving hadronic final states.  For example, jet substructure techniques have previously enabled the study of color flow between the final states in top decay \cite{Gallicchio:2010sw,Abazov:2011vh,Aad:2015lxa}, an effect that relies on detecting soft gluons from color-connected partons.  

In this paper, we have shown how to test the dead cone effect---a universal prediction of gauge theories---in QCD FSR from top quarks.  The top quark is rather special in this context, since for bottom and charm quarks, the dead cone effect is obscured both by heavy hadron decays and non-perturbative physics.  Focusing on top/anti-top pairs with a single-lepton final state, we presented a complete analysis strategy based on the observable $\Theta^2_S$ that can be used to test for the presence of the dead cone effect.  Our technique exploits the ability of the soft drop algorithm to reconstruct the angular pattern of the radiated gluon, despite the complications faced by several blurring effects including $b$-quark FSR.

There are four key steps to our procedure.  First, we use soft drop to define the candidate $b$ quark and gluon kinematics within a boosted leptonic top.  Second, we reconstruct the missing neutrino using the $W$ mass constraint.  Third, we select the $S$-enriched region of phase space where $m_{b\ell\nu} \simeq m_t$, such that the gluon is more likely to come from top FSR than from top decay.  Finally, we impose a $\Theta_b$ angular cut on the candidate $b$ quark to suppress residual contamination from $b$-quark FSR.  This leads to a subtle but convincing dead cone suppression in the $\Theta^2_S$ distribution.

In preliminary tests using a fast detector simulation \cite{deFavereau:2013fsa}, we find that the reconstructed top and gluon kinematics are not dramatically distorted by detector effects, owing partly to the robustness of the soft drop procedure.  We therefore look forward to detailed dead cone studies at the LHC, as well as future applications of jet substructure techniques to probe the subtleties of QCD.

\acknowledgments{We thank Torbj\"{o}rn Sj\"{o}strand for help validating the dead cone effect in \pythia{}. FM would like to thank Scott Willenbrock for inspiring this investigation. This work has been supported by the MIT-Belgium Program of the MIT International Science and Technology Initiatives.  The work of FM and MS has been performed in the framework of the ERC Grant No.  291377 ``LHCTheory'', it has been supported in part by the European Union as part of the FP7 Marie Curie Initial Training Network MCnetITN (PITN-GA-2012-315877), and in part by the Belgian Federal Science Policy Office through the Interuniversity Attraction Pole P7/37, and by the FNRS. The work of JT is supported by the U.S. Department of Energy (DOE) under cooperative research agreement DE-SC-00012567, by the DOE Early Career research program DE-SC-0006389, and by a Sloan Research Fellowship from the Alfred P. Sloan Foundation.}

\appendix

\section{Effect of final-state interference and bleed-through}
\label{app:interference}

\begin{figure}
\includegraphics[width=0.95\columnwidth]{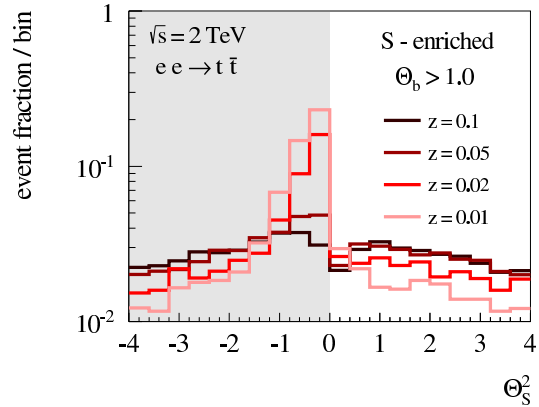}
\caption{The $\Theta_S^2$ distribution in $e^+ e^- \to t b \ell \nu g$ events at $2~\TeV$, for various choices of the minimum gluon energy.  The cuts $m_{b\ell\nu} \in [170,200]~\GeV$ and $\Theta_b > 1.0$ have been applied in order to reduce direct contamination from radiation in decay. For high enough gluon energies ($z \gtrsim 0.05$), the dead cone pattern is preserved whereas for smaller values ($z \lesssim 0.02$), it is washed out by interference and bleed-through effects.}
\label{fig:ee_int_z}
\end{figure}

As argued in \Eq{eq:width2}, a cut on the gluon energy fraction is necessary to avoid interference between the top FSR diagram ($S$) and the top decay diagrams ($B_{1,2}$) as well as to suppress bleed-through of the $B_{1,2}$ diagrams when $m_{b \ell \nu g} \approx m_t$.  To quantify this, we simulate $e^+ e^- \to t b \ell \nu g$ at tree level with \aNLO~in order to evaluate the full matrix element including the interference, where the same caveat from footnote \ref{footnote:thadsubtle} applies.

In \Fig{fig:ee_int_z}, we show the $\Theta_S^2$ observable obtained with a set of different cuts on the gluon energy.  Here, the final-state kinematics are assumed to be perfectly known, and we apply the same signal selection as the one described in \Secs{sec:sigiso}{sec:thetab}. Despite having explicitly suppressed direct contributions from $B_{1,2}$ by requiring $m_{b\ell\nu} \in [170,200]~\GeV$ and $\Theta_b > 1.0$, we see that for small gluon energies ($z \lesssim 0.02$), the dead cone is washed out by interference and bleed-through effects, confirming the qualitative arguments given in \Sec{sec:width}.

\section{Distributions without a $\Theta_b$ cut}
\label{app:thetab}

\begin{figure*}[t]
\subfloat[]{\label{fig:pp_tt_sg_1D}
\includegraphics[width=0.95\columnwidth]{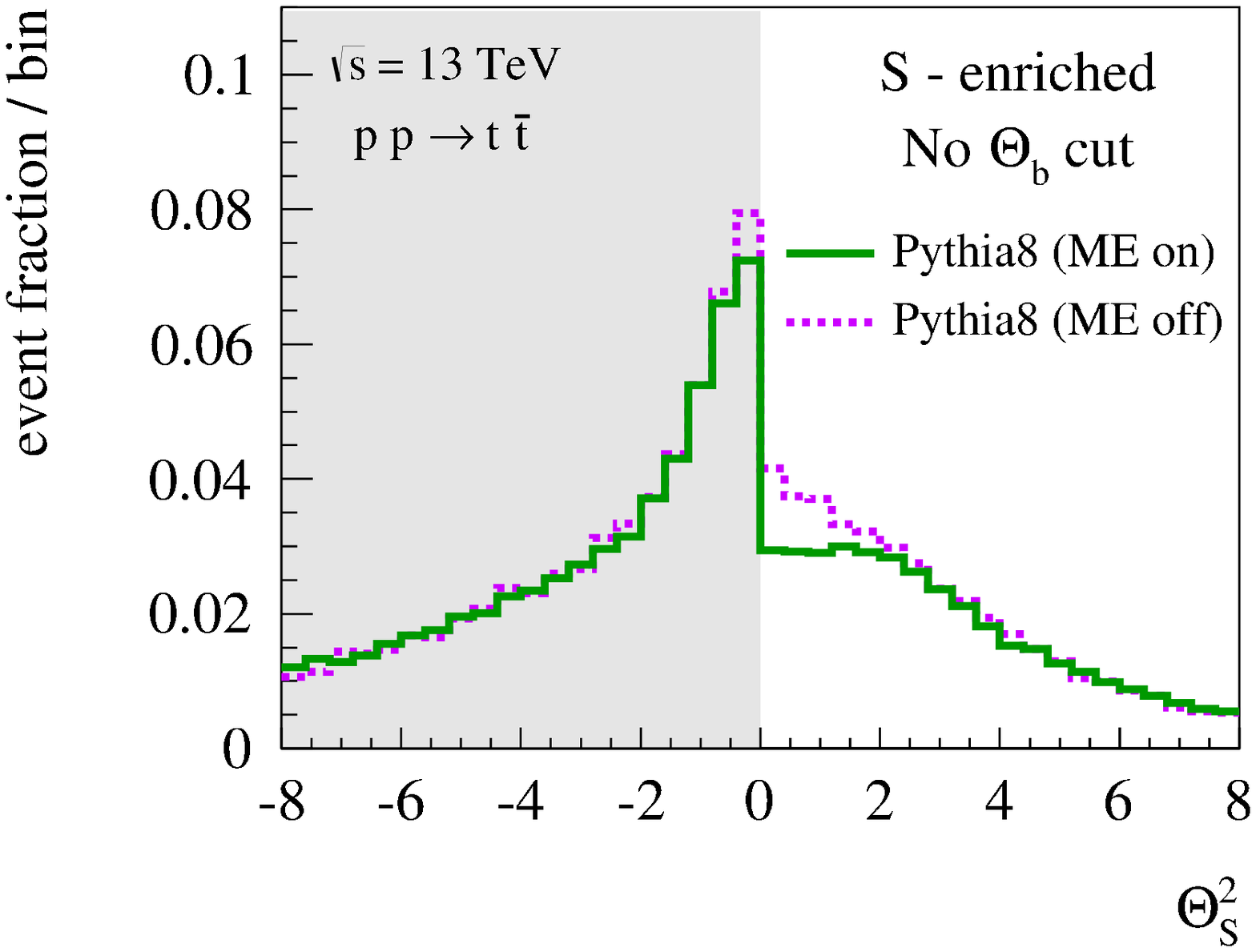}
}
$\quad$
\subfloat[]{\label{fig:pp_tt_bk_1D}
\includegraphics[width=0.95\columnwidth]{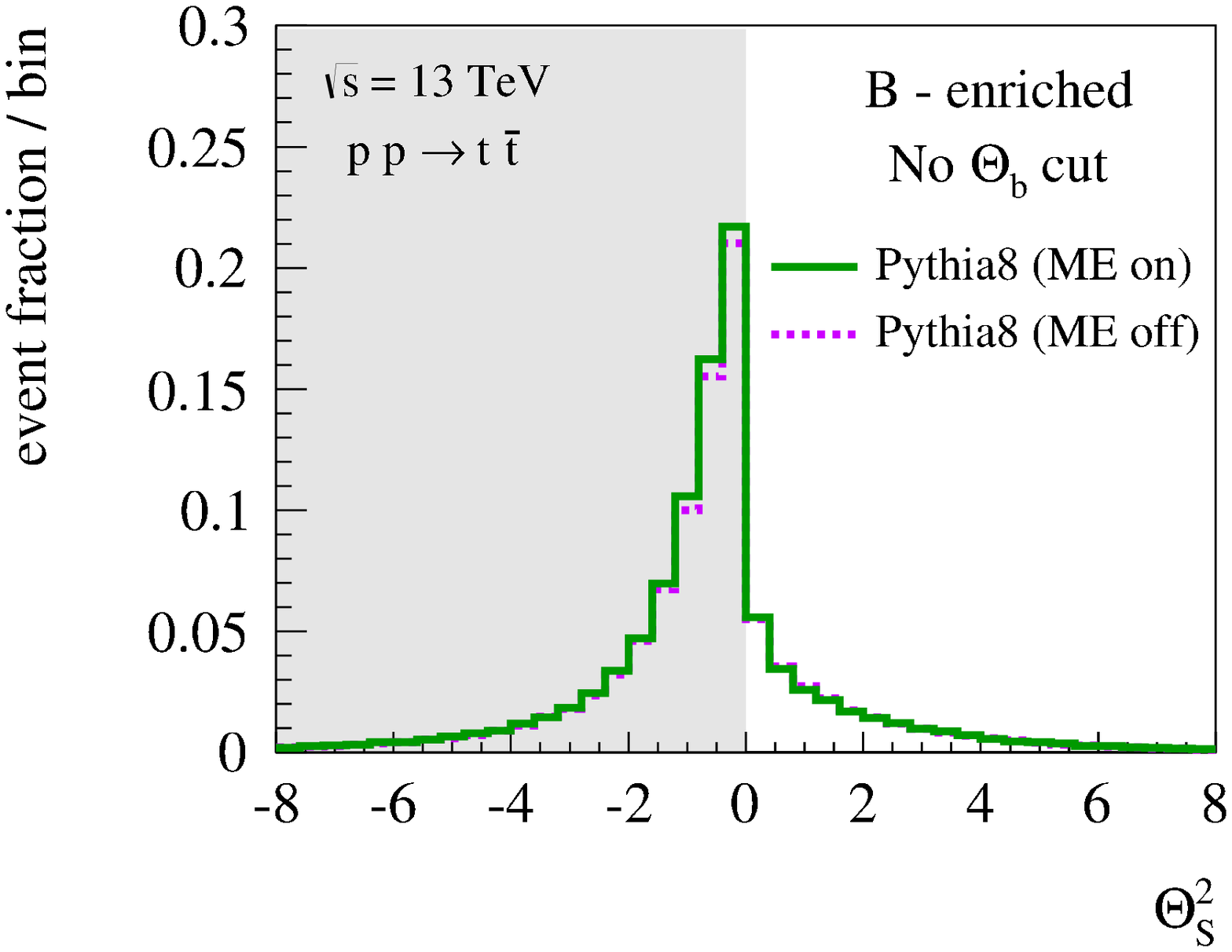}
}

\subfloat[]{\label{fig:pp_tt_sg_1D_ps}
\includegraphics[width=0.95\columnwidth]{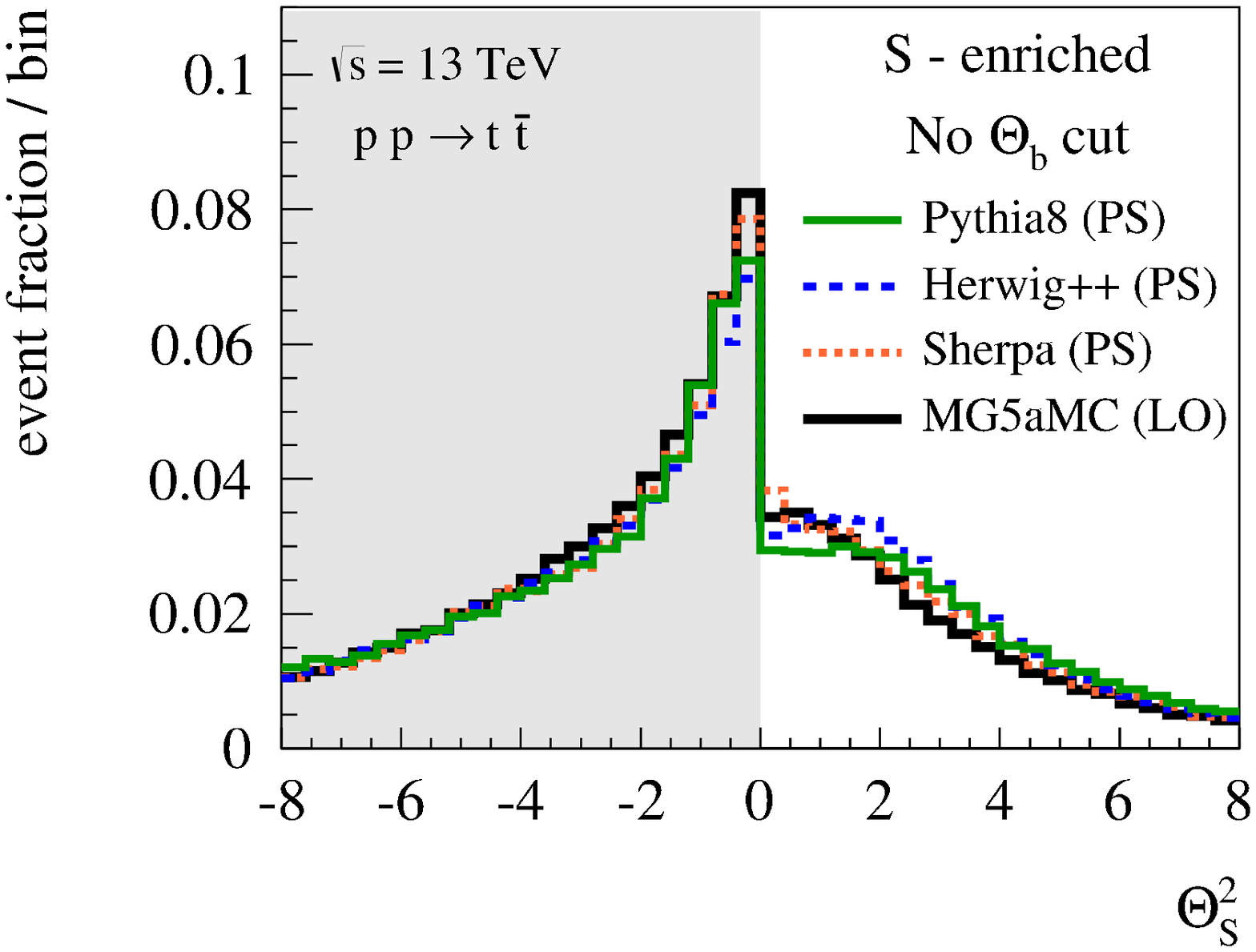}
}
$\quad$
\subfloat[]{\label{fig:pp_tt_bk_1D_ps}
\includegraphics[width=0.95\columnwidth]{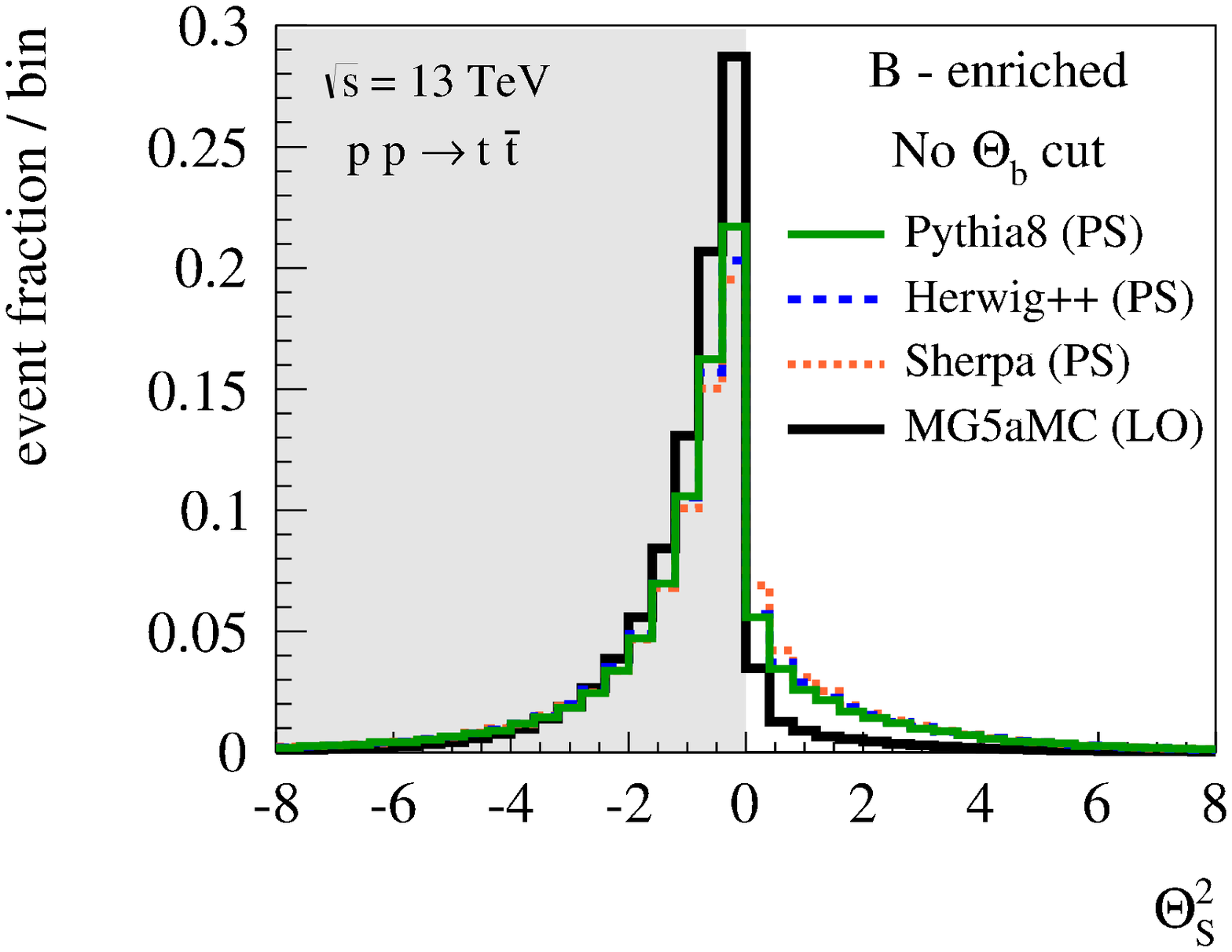}
}
\caption{The same as \Fig{fig:pp_tt_Th10}, but without a cut on $\Theta_b$.}
\label{fig:pp_tt}
\end{figure*}

Previously in \Fig{fig:pp_tt_Th10}, we showed distributions for $\Theta^2_S$ after imposing a cut of $\Theta_b > 1.0$.  The motivation for the $\Theta_b$ requirement was to increase the statistical significance of the dead cone effect.  Here in \Fig{fig:pp_tt}, we show results corresponding to the same analysis, yet without the $\Theta_b$ cut.  Comparing the \pythia{} distributions with ME corrections on and off, a statistically significant difference of roughly $3 \sigma$ after 300~fb$^{-1}$ can be observed.  The expected characteristic distribution of the radiation, however, is washed out, leading to a plateau between $0 < \Theta^2_S < 1$ instead of a suppression towards the origin.  We therefore conclude that a $\Theta_b$ cut will likely be needed to gain confidence in the dead cone effect.

\section{Idealized distributions with realistic cuts}
\label{app:eereal}

\begin{figure*}[t]
\subfloat[]{\label{fig:ee_tt_sg_2D_Th10}
\includegraphics[width=0.95\columnwidth]{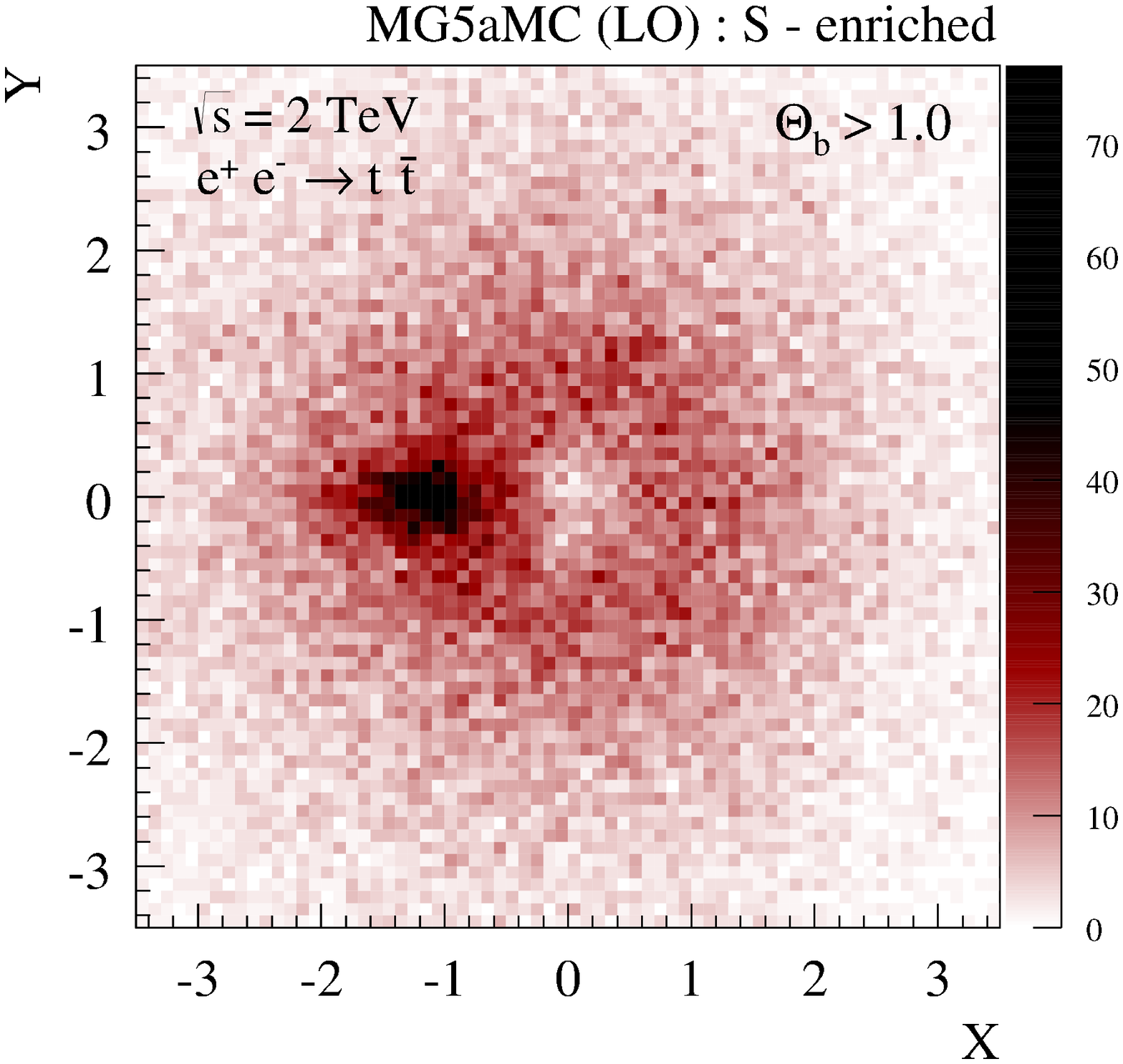}
}
$\quad$
\subfloat[]{\label{fig:ee_tt_bk_2D_Th10}
\includegraphics[width=0.95\columnwidth]{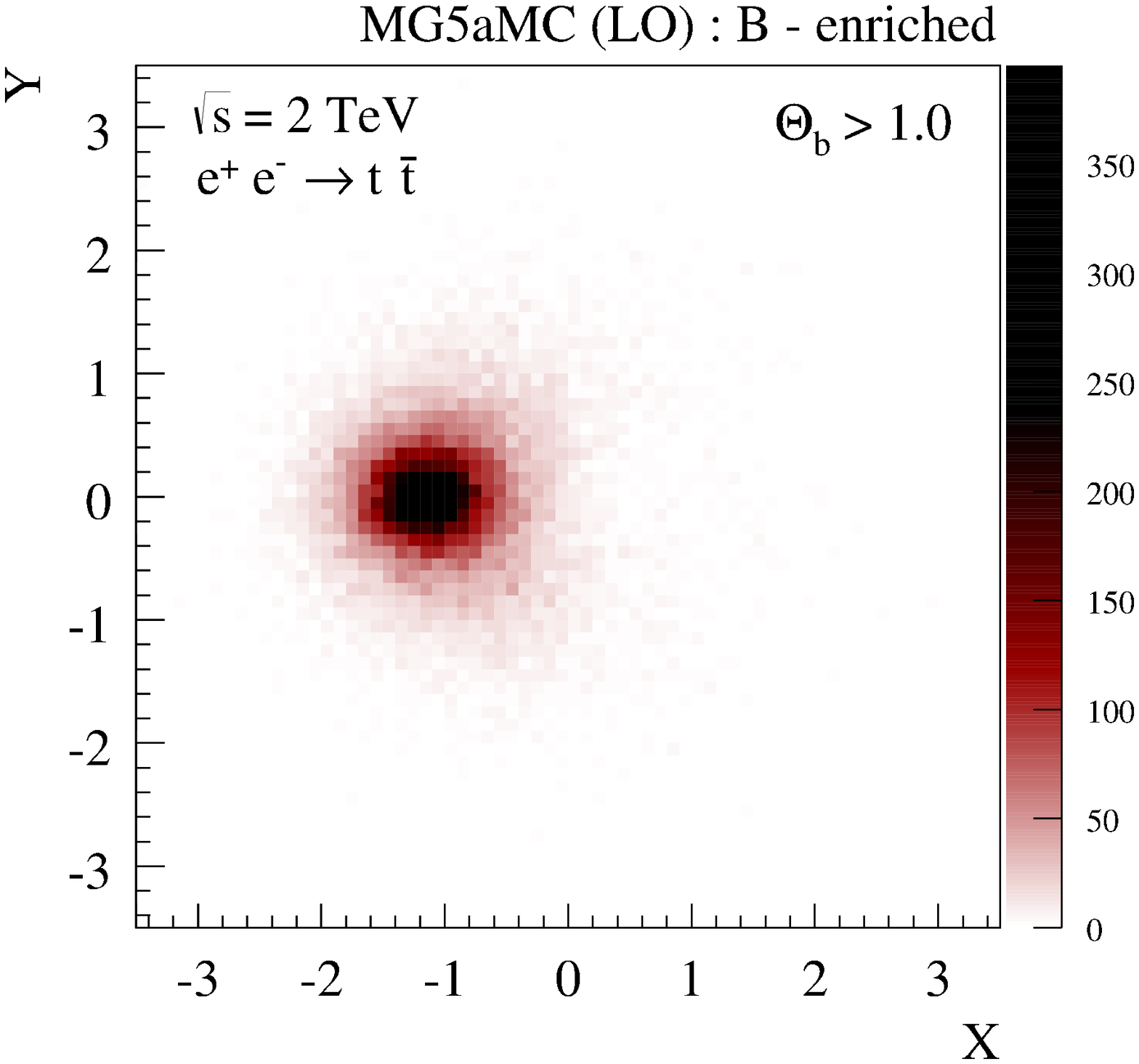}
}
\caption{The same as \Fig{fig:pp_tt_2D_Th10}, but applying our analysis strategy on $e^+ e^- \to t \bar{t}$ events at $\sqrt{s} = 2~\TeV$.}
\label{fig:ee_tt_2D_Th10}
\end{figure*}

As mentioned in \Sec{sec:thetab}, ISR and UE are sources of jet contamination that partially fill the dead cone in $pp$ collisions.  To understand the effect of this contamination on our analysis, we return to $e^+e^-$ collisions and use them as a template where ISR/UE effects are absent.  We then perform the exact same LHC-targeted analysis from \Sec{sec:strategy}, with the corresponding $e^+ e^-$ results shown in \Figs{fig:ee_tt_2D_Th10}{fig:ee_tt_Th10}.  The overall qualitative features are the same as in the $pp$ case.  As expected, though, the differences between the ME-on and ME-off \pythia{} distributions are more noticeable in the $e^+ e^-$ case, and the dip towards $\Theta_S^2 = 0$ is more pronounced.

\begin{figure*}[t]
\subfloat[]{\label{fig:ee_tt_sg_1D_Th10}
\includegraphics[width=0.95\columnwidth]{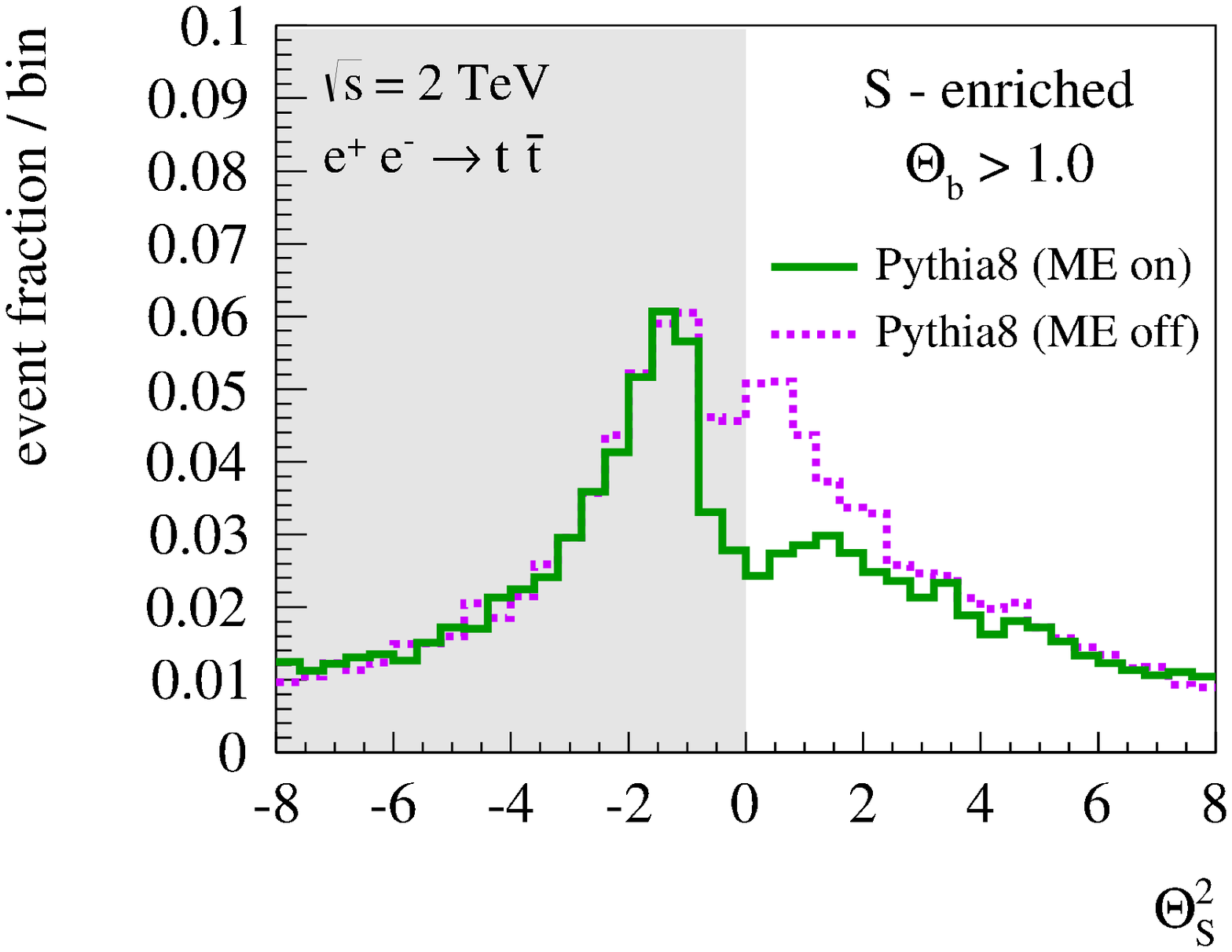}
}
$\quad$
\subfloat[]{\label{fig:ee_tt_bk_1D_Th10}
\includegraphics[width=0.95\columnwidth]{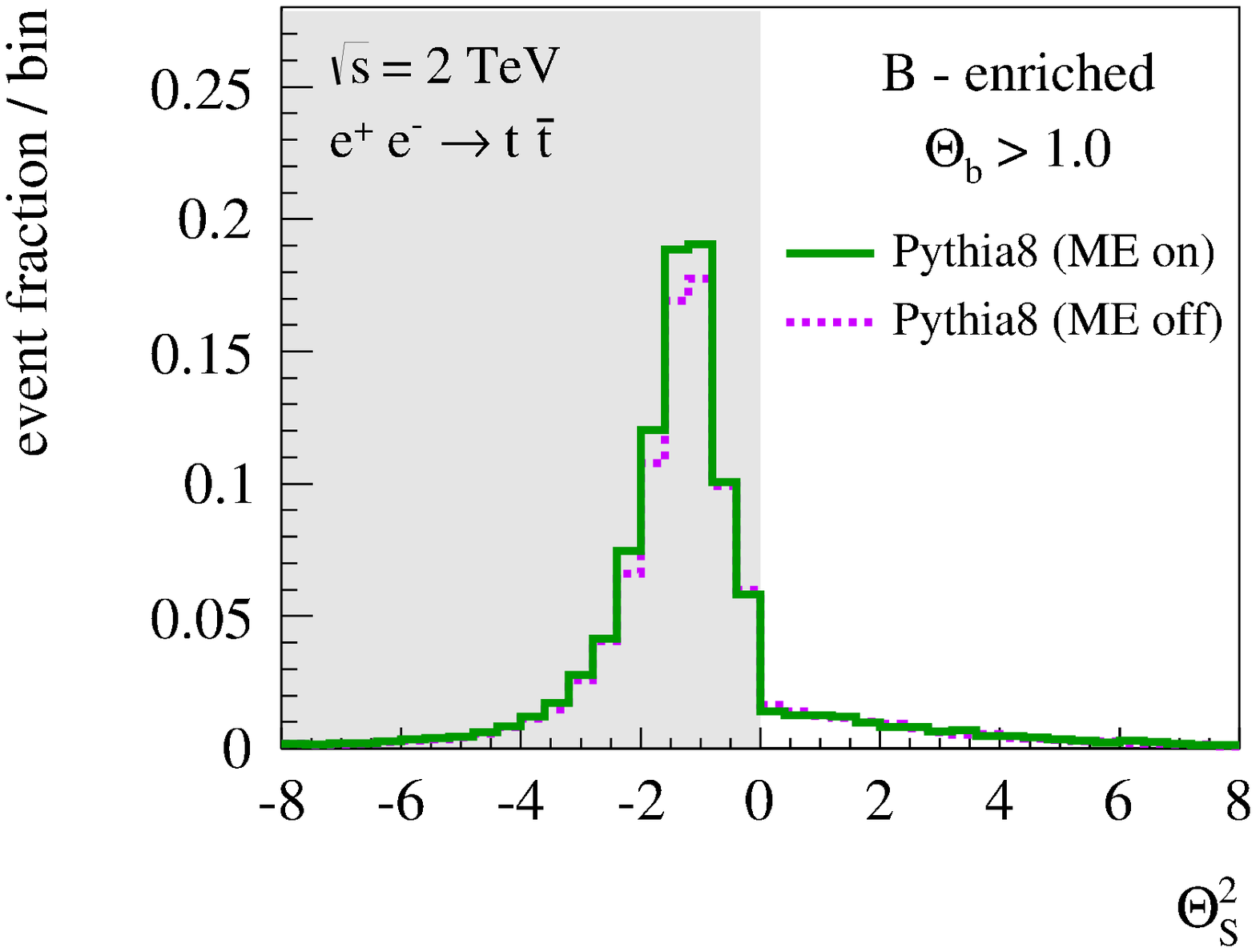}
}

\subfloat[]{\label{fig:ee_tt_sg_1D_ps_Th10}
\includegraphics[width=0.95\columnwidth]{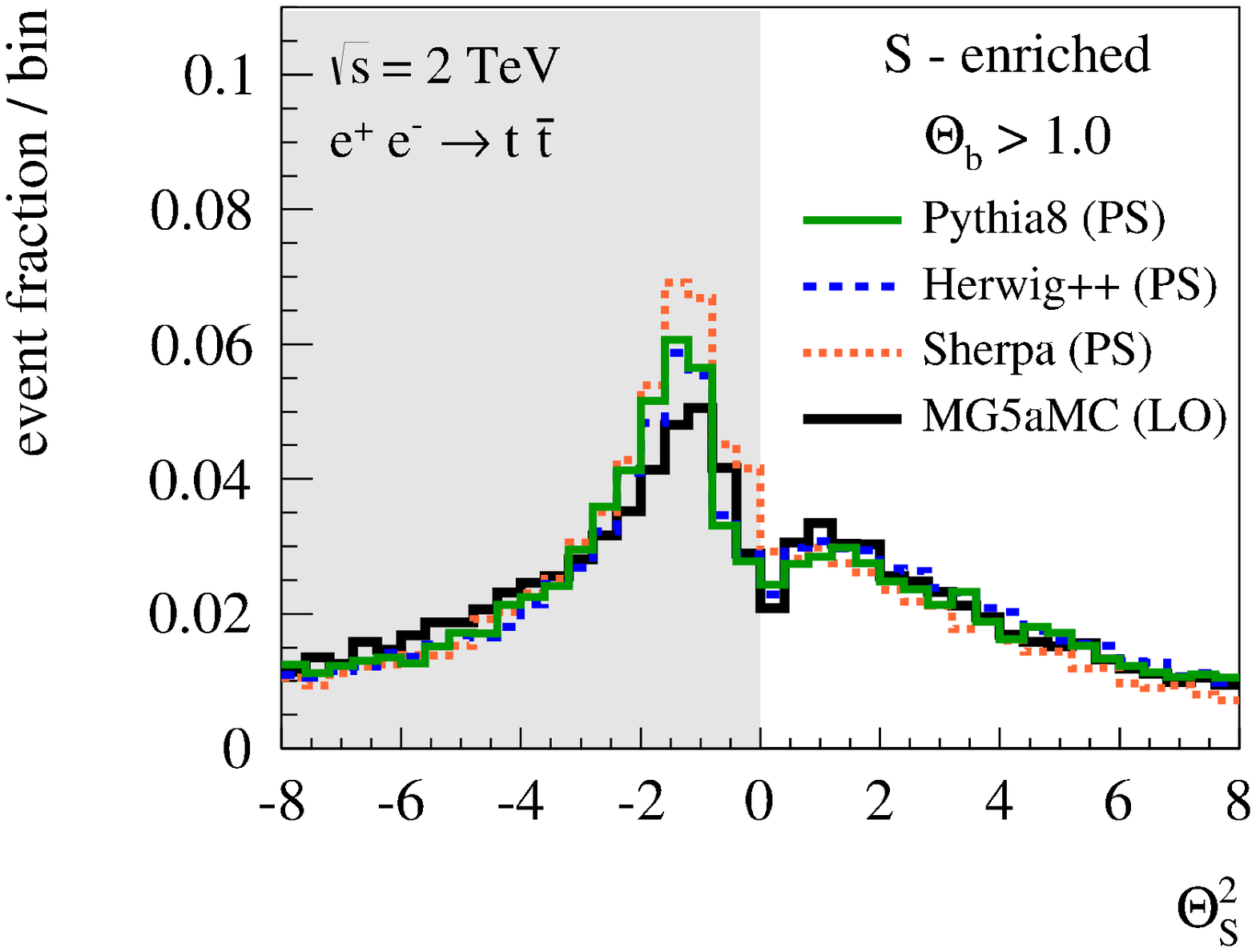}
}
$\quad$
\subfloat[]{\label{fig:ee_tt_bk_1D_ps_Th10}
\includegraphics[width=0.95\columnwidth]{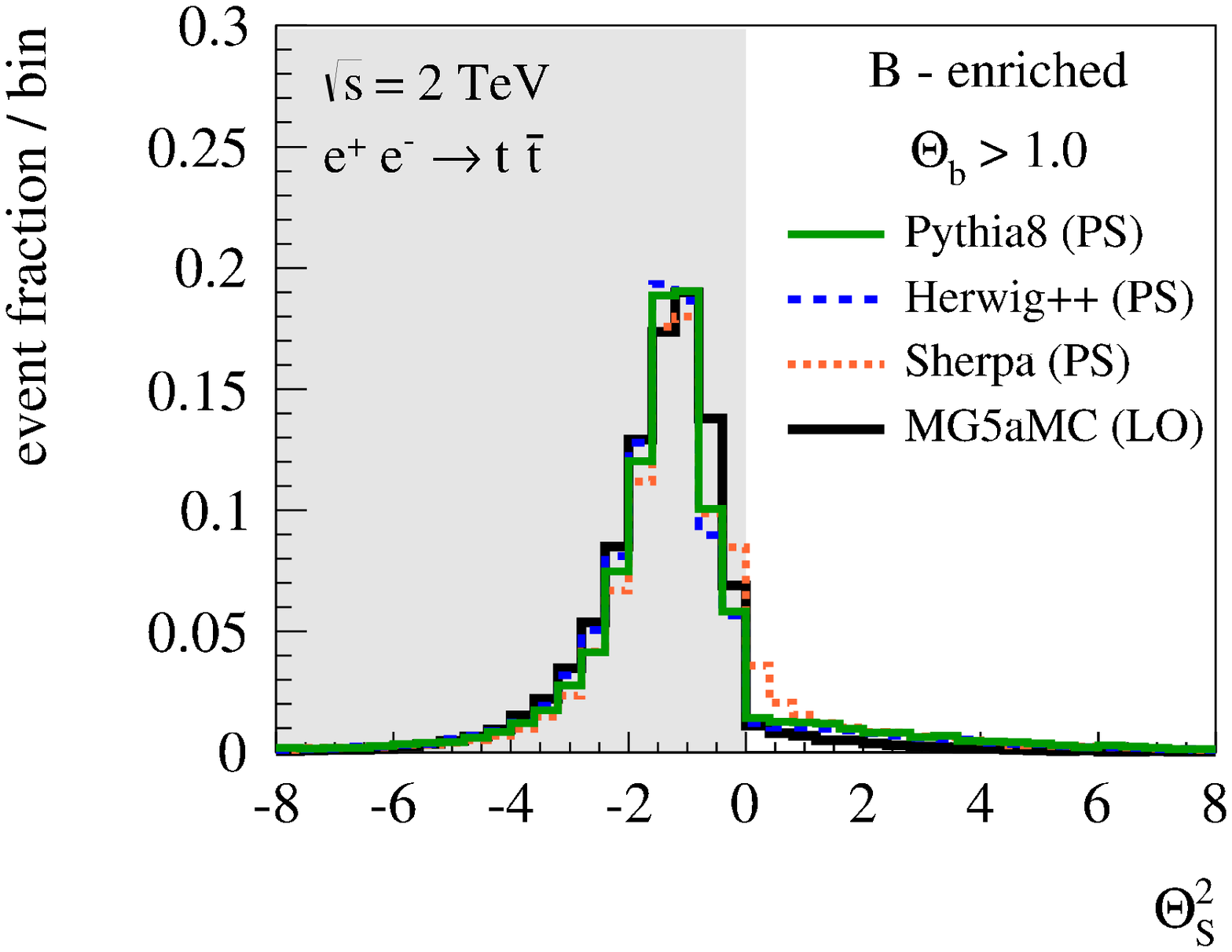}
}
\caption{The same as \Fig{fig:pp_tt_Th10}, but applying our analysis strategy on $e^+ e^- \to t \bar{t}$ events at $\sqrt{s} = 2~\TeV$.}
\label{fig:ee_tt_Th10}
\end{figure*}

\bibliography{deadcone}

\begin{thebibliography}{49}%
\makeatletter
\providecommand \@ifxundefined [1]{%
 \@ifx{#1\undefined}
}%
\providecommand \@ifnum [1]{%
 \ifnum #1\expandafter \@firstoftwo
 \else \expandafter \@secondoftwo
 \fi
}%
\providecommand \@ifx [1]{%
 \ifx #1\expandafter \@firstoftwo
 \else \expandafter \@secondoftwo
 \fi
}%
\providecommand \natexlab [1]{#1}%
\providecommand \enquote  [1]{``#1''}%
\providecommand \bibnamefont  [1]{#1}%
\providecommand \bibfnamefont [1]{#1}%
\providecommand \citenamefont [1]{#1}%
\providecommand \href@noop [0]{\@secondoftwo}%
\providecommand \href [0]{\begingroup \@sanitize@url \@href}%
\providecommand \@href[1]{\@@startlink{#1}\@@href}%
\providecommand \@@href[1]{\endgroup#1\@@endlink}%
\providecommand \@sanitize@url [0]{\catcode `\\12\catcode `\$12\catcode
  `\&12\catcode `\#12\catcode `\^12\catcode `\_12\catcode `\%12\relax}%
\providecommand \@@startlink[1]{}%
\providecommand \@@endlink[0]{}%
\providecommand \url  [0]{\begingroup\@sanitize@url \@url }%
\providecommand \@url [1]{\endgroup\@href {#1}{\urlprefix }}%
\providecommand \urlprefix  [0]{URL }%
\providecommand \Eprint [0]{\href }%
\providecommand \doibase [0]{http://dx.doi.org/}%
\providecommand \selectlanguage [0]{\@gobble}%
\providecommand \bibinfo  [0]{\@secondoftwo}%
\providecommand \bibfield  [0]{\@secondoftwo}%
\providecommand \translation [1]{[#1]}%
\providecommand \BibitemOpen [0]{}%
\providecommand \bibitemStop [0]{}%
\providecommand \bibitemNoStop [0]{.\EOS\space}%
\providecommand \EOS [0]{\spacefactor3000\relax}%
\providecommand \BibitemShut  [1]{\csname bibitem#1\endcsname}%
\let\auto@bib@innerbib\@empty
\bibitem [{\citenamefont {Dokshitzer}\ \emph
  {et~al.}(1991{\natexlab{a}})\citenamefont {Dokshitzer}, \citenamefont
  {Khoze},\ and\ \citenamefont {Troian}}]{Dokshitzer:1991fc}%
  \BibitemOpen
  \bibfield  {author} {\bibinfo {author} {\bibfnamefont {Yuri~L.}\ \bibnamefont
  {Dokshitzer}}, \bibinfo {author} {\bibfnamefont {Valery~A.}\ \bibnamefont
  {Khoze}}, \ and\ \bibinfo {author} {\bibfnamefont {S.~I.}\ \bibnamefont
  {Troian}},\ }\bibfield  {title} {\enquote {\bibinfo {title} {{Particle
  spectra in light and heavy quark jets}},}\ }\bibfield  {booktitle} {\emph
  {\bibinfo {booktitle} {{Jet Studies Workshop at LEP and HERA Durham, England,
  December 9-15, 1990}}},\ }\href {\doibase 10.1088/0954-3899/17/10/003}
  {\bibfield  {journal} {\bibinfo  {journal} {J. Phys.}\ }\textbf {\bibinfo
  {volume} {G17}},\ \bibinfo {pages} {1481--1492} (\bibinfo {year}
  {1991}{\natexlab{a}})}\BibitemShut {NoStop}%
\bibitem [{\citenamefont {Dokshitzer}\ \emph
  {et~al.}(1991{\natexlab{b}})\citenamefont {Dokshitzer}, \citenamefont
  {Khoze},\ and\ \citenamefont {Troian}}]{Dokshitzer:1991fd}%
  \BibitemOpen
  \bibfield  {author} {\bibinfo {author} {\bibfnamefont {Yuri~L.}\ \bibnamefont
  {Dokshitzer}}, \bibinfo {author} {\bibfnamefont {Valery~A.}\ \bibnamefont
  {Khoze}}, \ and\ \bibinfo {author} {\bibfnamefont {S.~I.}\ \bibnamefont
  {Troian}},\ }\bibfield  {title} {\enquote {\bibinfo {title} {{On specific QCD
  properties of heavy quark fragmentation ('dead cone')}},}\ }\bibfield
  {booktitle} {\emph {\bibinfo {booktitle} {{Jet Studies Workshop at LEP and
  HERA Durham, England, December 9-15, 1990}}},\ }\href {\doibase
  10.1088/0954-3899/17/10/023} {\bibfield  {journal} {\bibinfo  {journal} {J.
  Phys.}\ }\textbf {\bibinfo {volume} {G17}},\ \bibinfo {pages} {1602--1604}
  (\bibinfo {year} {1991}{\natexlab{b}})}\BibitemShut {NoStop}%
\bibitem [{\citenamefont {Ellis}\ \emph {et~al.}(1996)\citenamefont {Ellis},
  \citenamefont {Stirling},\ and\ \citenamefont {Webber}}]{Ellis:1991qj}%
  \BibitemOpen
  \bibfield  {author} {\bibinfo {author} {\bibfnamefont {R.~Keith}\
  \bibnamefont {Ellis}}, \bibinfo {author} {\bibfnamefont {W.~James}\
  \bibnamefont {Stirling}}, \ and\ \bibinfo {author} {\bibfnamefont {B.~R.}\
  \bibnamefont {Webber}},\ }\bibfield  {title} {\enquote {\bibinfo {title}
  {{QCD and collider physics}},}\ }\href@noop {} {\bibfield  {journal}
  {\bibinfo  {journal} {Camb. Monogr. Part. Phys. Nucl. Phys. Cosmol.}\
  }\textbf {\bibinfo {volume} {8}},\ \bibinfo {pages} {1--435} (\bibinfo {year}
  {1996})}\BibitemShut {NoStop}%
\bibitem [{\citenamefont {Schumm}\ \emph {et~al.}(1992)\citenamefont {Schumm},
  \citenamefont {Dokshitzer}, \citenamefont {Khoze},\ and\ \citenamefont
  {Koetke}}]{Schumm:1992xt}%
  \BibitemOpen
  \bibfield  {author} {\bibinfo {author} {\bibfnamefont {Bruce~A.}\
  \bibnamefont {Schumm}}, \bibinfo {author} {\bibfnamefont {Yuri~L.}\
  \bibnamefont {Dokshitzer}}, \bibinfo {author} {\bibfnamefont {Valery~A.}\
  \bibnamefont {Khoze}}, \ and\ \bibinfo {author} {\bibfnamefont {Dale~S.}\
  \bibnamefont {Koetke}},\ }\bibfield  {title} {\enquote {\bibinfo {title}
  {{MLLA and the average charged multiplicity of events containing heavy quarks
  in e+ e- annihilation}},}\ }\href {\doibase 10.1103/PhysRevLett.69.3025}
  {\bibfield  {journal} {\bibinfo  {journal} {Phys. Rev. Lett.}\ }\textbf
  {\bibinfo {volume} {69}},\ \bibinfo {pages} {3025--3028} (\bibinfo {year}
  {1992})}\BibitemShut {NoStop}%
\bibitem [{\citenamefont {Dokshitzer}\ \emph {et~al.}(1996)\citenamefont
  {Dokshitzer}, \citenamefont {Khoze},\ and\ \citenamefont
  {Troian}}]{Dokshitzer:1995ev}%
  \BibitemOpen
  \bibfield  {author} {\bibinfo {author} {\bibfnamefont {Yuri~L.}\ \bibnamefont
  {Dokshitzer}}, \bibinfo {author} {\bibfnamefont {Valery~A.}\ \bibnamefont
  {Khoze}}, \ and\ \bibinfo {author} {\bibfnamefont {S.~I.}\ \bibnamefont
  {Troian}},\ }\bibfield  {title} {\enquote {\bibinfo {title} {{Specific
  features of heavy quark production. LPHD approach to heavy particle
  spectra}},}\ }\href {\doibase 10.1103/PhysRevD.53.89} {\bibfield  {journal}
  {\bibinfo  {journal} {Phys. Rev.}\ }\textbf {\bibinfo {volume} {D53}},\
  \bibinfo {pages} {89--119} (\bibinfo {year} {1996})},\ \Eprint
  {http://arxiv.org/abs/hep-ph/9506425} {arXiv:hep-ph/9506425 [hep-ph]}
  \BibitemShut {NoStop}%
\bibitem [{\citenamefont {Dokshitzer}\ \emph {et~al.}(2006)\citenamefont
  {Dokshitzer}, \citenamefont {Fabbri}, \citenamefont {Khoze},\ and\
  \citenamefont {Ochs}}]{Dokshitzer:2005ri}%
  \BibitemOpen
  \bibfield  {author} {\bibinfo {author} {\bibfnamefont {Yuri~L.}\ \bibnamefont
  {Dokshitzer}}, \bibinfo {author} {\bibfnamefont {Fabrizio}\ \bibnamefont
  {Fabbri}}, \bibinfo {author} {\bibfnamefont {Valery~A.}\ \bibnamefont
  {Khoze}}, \ and\ \bibinfo {author} {\bibfnamefont {Wolfgang}\ \bibnamefont
  {Ochs}},\ }\bibfield  {title} {\enquote {\bibinfo {title} {{Multiplicity
  difference between heavy and light quark jets revisited}},}\ }\href {\doibase
  10.1140/epjc/s2005-02424-5} {\bibfield  {journal} {\bibinfo  {journal} {Eur.
  Phys. J.}\ }\textbf {\bibinfo {volume} {C45}},\ \bibinfo {pages} {387--400}
  (\bibinfo {year} {2006})},\ \Eprint {http://arxiv.org/abs/hep-ph/0508074}
  {arXiv:hep-ph/0508074 [hep-ph]} \BibitemShut {NoStop}%
\bibitem [{\citenamefont {Perieanu}(2006)}]{Perieanu:2006vn}%
  \BibitemOpen
  \bibfield  {author} {\bibinfo {author} {\bibfnamefont {Adrian}\ \bibnamefont
  {Perieanu}},\ }\emph {\bibinfo {title} {{The Structure of Charm Jets and the
  Dead Cone Effect in Deep-Inelastic Scattering at HERA}}},\ \href {\doibase
  10.3204/DESY-THESIS-2006-002} {Ph.D. thesis},\ \bibinfo  {school} {Hamburg
  U.} (\bibinfo {year} {2006})\BibitemShut {NoStop}%
\bibitem [{\citenamefont {Abdallah}\ \emph {et~al.}(2008)\citenamefont
  {Abdallah} \emph {et~al.}}]{Abdallah:2008ac}%
  \BibitemOpen
  \bibfield  {author} {\bibinfo {author} {\bibfnamefont {J.}~\bibnamefont
  {Abdallah}} \emph {et~al.} (\bibinfo {collaboration} {DELPHI}),\ }\bibfield
  {title} {\enquote {\bibinfo {title} {{Study of b-quark mass effects in
  multijet topologies with the DELPHI detector at LEP}},}\ }\href {\doibase
  10.1140/epjc/s10052-008-0631-5} {\bibfield  {journal} {\bibinfo  {journal}
  {Eur. Phys. J.}\ }\textbf {\bibinfo {volume} {C55}},\ \bibinfo {pages}
  {525--538} (\bibinfo {year} {2008})},\ \Eprint
  {http://arxiv.org/abs/0804.3883} {arXiv:0804.3883 [hep-ex]} \BibitemShut
  {NoStop}%
\bibitem [{\citenamefont {Marchesini}\ and\ \citenamefont
  {Webber}(1990)}]{Marchesini:1989yk}%
  \BibitemOpen
  \bibfield  {author} {\bibinfo {author} {\bibfnamefont {G.}~\bibnamefont
  {Marchesini}}\ and\ \bibinfo {author} {\bibfnamefont {B.~R.}\ \bibnamefont
  {Webber}},\ }\bibfield  {title} {\enquote {\bibinfo {title} {{Simulation of
  {QCD} Coherence in Heavy Quark Production and Decay}},}\ }\href {\doibase
  10.1016/0550-3213(90)90310-A} {\bibfield  {journal} {\bibinfo  {journal}
  {Nucl. Phys.}\ }\textbf {\bibinfo {volume} {B330}},\ \bibinfo {pages}
  {261--283} (\bibinfo {year} {1990})}\BibitemShut {NoStop}%
\bibitem [{\citenamefont {Okabe}\ \emph {et~al.}(1998)\citenamefont {Okabe}
  \emph {et~al.}}]{Okabe:1997uf}%
  \BibitemOpen
  \bibfield  {author} {\bibinfo {author} {\bibfnamefont {K.}~\bibnamefont
  {Okabe}} \emph {et~al.} (\bibinfo {collaboration} {VENUS}),\ }\bibfield
  {title} {\enquote {\bibinfo {title} {{Measurement of the charged multiplicity
  of bottom and light quark events in e+ e- annihilation at s**(1/2) =
  58-GeV}},}\ }\href {\doibase 10.1016/S0370-2693(98)00073-2} {\bibfield
  {journal} {\bibinfo  {journal} {Phys. Lett.}\ }\textbf {\bibinfo {volume}
  {B423}},\ \bibinfo {pages} {407--418} (\bibinfo {year} {1998})}\BibitemShut
  {NoStop}%
\bibitem [{\citenamefont {Ramos}\ \emph {et~al.}(2010)\citenamefont {Ramos},
  \citenamefont {Mathieu},\ and\ \citenamefont
  {Sanchis-Lozano}}]{Ramos:2010cma}%
  \BibitemOpen
  \bibfield  {author} {\bibinfo {author} {\bibfnamefont {Redamy~Perez}\
  \bibnamefont {Ramos}}, \bibinfo {author} {\bibfnamefont {Vincent}\
  \bibnamefont {Mathieu}}, \ and\ \bibinfo {author} {\bibfnamefont
  {Miguel-Angel}\ \bibnamefont {Sanchis-Lozano}},\ }\bibfield  {title}
  {\enquote {\bibinfo {title} {{Heavy quark flavour dependence of multiparticle
  production in QCD jets}},}\ }\href {\doibase 10.1007/JHEP08(2010)047}
  {\bibfield  {journal} {\bibinfo  {journal} {JHEP}\ }\textbf {\bibinfo
  {volume} {08}},\ \bibinfo {pages} {047} (\bibinfo {year} {2010})},\ \Eprint
  {http://arxiv.org/abs/1005.1582} {arXiv:1005.1582 [hep-ph]} \BibitemShut
  {NoStop}%
\bibitem [{\citenamefont {Barlow}\ \emph {et~al.}(1991)\citenamefont {Barlow}
  \emph {et~al.}}]{Barlow:1991fe}%
  \BibitemOpen
  \bibfield  {author} {\bibinfo {author} {\bibfnamefont {Roger~J.}\
  \bibnamefont {Barlow}} \emph {et~al.},\ }\bibfield  {title} {\enquote
  {\bibinfo {title} {{Report of the heavy flavors working group}},}\ }\bibfield
   {booktitle} {\emph {\bibinfo {booktitle} {{Jet Studies Workshop at LEP and
  HERA Durham, England, December 9-15, 1990}}},\ }\href {\doibase
  10.1088/0954-3899/17/10/024} {\bibfield  {journal} {\bibinfo  {journal} {J.
  Phys.}\ }\textbf {\bibinfo {volume} {G17}},\ \bibinfo {pages} {1605--1623}
  (\bibinfo {year} {1991})}\BibitemShut {NoStop}%
\bibitem [{\citenamefont {Battaglia}\ \emph {et~al.}(2004)\citenamefont
  {Battaglia}, \citenamefont {Orava},\ and\ \citenamefont
  {Salmi}}]{Battaglia:989441}%
  \BibitemOpen
  \bibfield  {author} {\bibinfo {author} {\bibfnamefont {M}~\bibnamefont
  {Battaglia}}, \bibinfo {author} {\bibfnamefont {R}~\bibnamefont {Orava}}, \
  and\ \bibinfo {author} {\bibfnamefont {L}~\bibnamefont {Salmi}},\ }\href@noop
  {} {\emph {\bibinfo {title} {{A Study of depletion of fragmentation particles
  at small angles in b-jets with the DELPHI detector at LEP}}}},\ \bibinfo
  {type} {Tech. Rep.}\ \bibinfo {number} {DELPHI-2004-037-CONF-712.
  CERN-DELPHI-2004-037-CONF-712}\ (\bibinfo  {institution} {CERN},\ \bibinfo
  {address} {Geneva},\ \bibinfo {year} {2004})\BibitemShut {NoStop}%
\bibitem [{\citenamefont {Chekanov}(2000)}]{Chekanov:2000et}%
  \BibitemOpen
  \bibfield  {author} {\bibinfo {author} {\bibfnamefont {S.~V.}\ \bibnamefont
  {Chekanov}},\ }\bibfield  {title} {\enquote {\bibinfo {title} {{Soft gluon
  angular screening in heavy quark fragmentation}},}\ }\href {\doibase
  10.1016/S0370-2693(00)00619-5} {\bibfield  {journal} {\bibinfo  {journal}
  {Phys. Lett.}\ }\textbf {\bibinfo {volume} {B484}},\ \bibinfo {pages}
  {51--57} (\bibinfo {year} {2000})},\ \Eprint
  {http://arxiv.org/abs/hep-ph/0005119} {arXiv:hep-ph/0005119 [hep-ph]}
  \BibitemShut {NoStop}%
\bibitem [{\citenamefont {Soper}\ and\ \citenamefont
  {Spannowsky}(2013)}]{Soper:2012pb}%
  \BibitemOpen
  \bibfield  {author} {\bibinfo {author} {\bibfnamefont {Davison~E.}\
  \bibnamefont {Soper}}\ and\ \bibinfo {author} {\bibfnamefont {Michael}\
  \bibnamefont {Spannowsky}},\ }\bibfield  {title} {\enquote {\bibinfo {title}
  {{Finding top quarks with shower deconstruction}},}\ }\href {\doibase
  10.1103/PhysRevD.87.054012} {\bibfield  {journal} {\bibinfo  {journal} {Phys.
  Rev.}\ }\textbf {\bibinfo {volume} {D87}},\ \bibinfo {pages} {054012}
  (\bibinfo {year} {2013})},\ \Eprint {http://arxiv.org/abs/1211.3140}
  {arXiv:1211.3140 [hep-ph]} \BibitemShut {NoStop}%
\bibitem [{\citenamefont {Abdesselam}\ \emph {et~al.}(2011)\citenamefont
  {Abdesselam}, \citenamefont {Kuutmann}, \citenamefont {Bitenc}, \citenamefont
  {Brooijmans}, \citenamefont {Butterworth} \emph
  {et~al.}}]{Abdesselam:2010pt}%
  \BibitemOpen
  \bibfield  {author} {\bibinfo {author} {\bibfnamefont {A.}~\bibnamefont
  {Abdesselam}}, \bibinfo {author} {\bibfnamefont {E.~Bergeaas}\ \bibnamefont
  {Kuutmann}}, \bibinfo {author} {\bibfnamefont {U.}~\bibnamefont {Bitenc}},
  \bibinfo {author} {\bibfnamefont {G.}~\bibnamefont {Brooijmans}}, \bibinfo
  {author} {\bibfnamefont {J.}~\bibnamefont {Butterworth}},  \emph {et~al.},\
  }\bibfield  {title} {\enquote {\bibinfo {title} {{Boosted objects: A Probe of
  beyond the Standard Model physics}},}\ }\href {\doibase
  10.1140/epjc/s10052-011-1661-y} {\bibfield  {journal} {\bibinfo  {journal}
  {Eur.Phys.J.}\ }\textbf {\bibinfo {volume} {C71}},\ \bibinfo {pages} {1661}
  (\bibinfo {year} {2011})},\ \Eprint {http://arxiv.org/abs/1012.5412}
  {arXiv:1012.5412 [hep-ph]} \BibitemShut {NoStop}%
\bibitem [{\citenamefont {Altheimer}\ \emph {et~al.}(2012)\citenamefont
  {Altheimer}, \citenamefont {Arora}, \citenamefont {Asquith}, \citenamefont
  {Brooijmans}, \citenamefont {Butterworth} \emph {et~al.}}]{Altheimer:2012mn}%
  \BibitemOpen
  \bibfield  {author} {\bibinfo {author} {\bibfnamefont {A.}~\bibnamefont
  {Altheimer}}, \bibinfo {author} {\bibfnamefont {S.}~\bibnamefont {Arora}},
  \bibinfo {author} {\bibfnamefont {L.}~\bibnamefont {Asquith}}, \bibinfo
  {author} {\bibfnamefont {G.}~\bibnamefont {Brooijmans}}, \bibinfo {author}
  {\bibfnamefont {J.}~\bibnamefont {Butterworth}},  \emph {et~al.},\ }\bibfield
   {title} {\enquote {\bibinfo {title} {{Jet Substructure at the Tevatron and
  LHC: New results, new tools, new benchmarks}},}\ }\href {\doibase
  10.1088/0954-3899/39/6/063001} {\bibfield  {journal} {\bibinfo  {journal}
  {J.Phys.}\ }\textbf {\bibinfo {volume} {G39}},\ \bibinfo {pages} {063001}
  (\bibinfo {year} {2012})},\ \Eprint {http://arxiv.org/abs/1201.0008}
  {arXiv:1201.0008 [hep-ph]} \BibitemShut {NoStop}%
\bibitem [{\citenamefont {Altheimer}\ \emph {et~al.}(2014)\citenamefont
  {Altheimer}, \citenamefont {Arce}, \citenamefont {Asquith}, \citenamefont
  {Backus~Mayes}, \citenamefont {Bergeaas~Kuutmann} \emph
  {et~al.}}]{Altheimer:2013yza}%
  \BibitemOpen
  \bibfield  {author} {\bibinfo {author} {\bibfnamefont {A.}~\bibnamefont
  {Altheimer}}, \bibinfo {author} {\bibfnamefont {A.}~\bibnamefont {Arce}},
  \bibinfo {author} {\bibfnamefont {L.}~\bibnamefont {Asquith}}, \bibinfo
  {author} {\bibfnamefont {J.}~\bibnamefont {Backus~Mayes}}, \bibinfo {author}
  {\bibfnamefont {E.}~\bibnamefont {Bergeaas~Kuutmann}},  \emph {et~al.},\
  }\bibfield  {title} {\enquote {\bibinfo {title} {{Boosted objects and jet
  substructure at the LHC. Report of BOOST2012, held at IFIC Valencia,
  23rd-27th of July 2012}},}\ }\href {\doibase 10.1140/epjc/s10052-014-2792-8}
  {\bibfield  {journal} {\bibinfo  {journal} {Eur.Phys.J.}\ }\textbf {\bibinfo
  {volume} {C74}},\ \bibinfo {pages} {2792} (\bibinfo {year} {2014})},\ \Eprint
  {http://arxiv.org/abs/1311.2708} {arXiv:1311.2708 [hep-ex]} \BibitemShut
  {NoStop}%
\bibitem [{\citenamefont {Adams}\ \emph {et~al.}(2015)\citenamefont {Adams}
  \emph {et~al.}}]{Adams:2015hiv}%
  \BibitemOpen
  \bibfield  {author} {\bibinfo {author} {\bibfnamefont {D.}~\bibnamefont
  {Adams}} \emph {et~al.},\ }\bibfield  {title} {\enquote {\bibinfo {title}
  {{Towards an Understanding of the Correlations in Jet Substructure}},}\
  }\href {\doibase 10.1140/epjc/s10052-015-3587-2} {\bibfield  {journal}
  {\bibinfo  {journal} {Eur. Phys. J.}\ }\textbf {\bibinfo {volume} {C75}},\
  \bibinfo {pages} {409} (\bibinfo {year} {2015})},\ \Eprint
  {http://arxiv.org/abs/1504.00679} {arXiv:1504.00679 [hep-ph]} \BibitemShut
  {NoStop}%
\bibitem [{\citenamefont {Larkoski}\ \emph {et~al.}(2014)\citenamefont
  {Larkoski}, \citenamefont {Marzani}, \citenamefont {Soyez},\ and\
  \citenamefont {Thaler}}]{Larkoski:2014wba}%
  \BibitemOpen
  \bibfield  {author} {\bibinfo {author} {\bibfnamefont {Andrew~J.}\
  \bibnamefont {Larkoski}}, \bibinfo {author} {\bibfnamefont {Simone}\
  \bibnamefont {Marzani}}, \bibinfo {author} {\bibfnamefont {Gregory}\
  \bibnamefont {Soyez}}, \ and\ \bibinfo {author} {\bibfnamefont {Jesse}\
  \bibnamefont {Thaler}},\ }\bibfield  {title} {\enquote {\bibinfo {title}
  {{Soft Drop}},}\ }\href {\doibase 10.1007/JHEP05(2014)146} {\bibfield
  {journal} {\bibinfo  {journal} {JHEP}\ }\textbf {\bibinfo {volume} {1405}},\
  \bibinfo {pages} {146} (\bibinfo {year} {2014})},\ \Eprint
  {http://arxiv.org/abs/1402.2657} {arXiv:1402.2657 [hep-ph]} \BibitemShut
  {NoStop}%
\bibitem [{\citenamefont {Butterworth}\ \emph {et~al.}(2008)\citenamefont
  {Butterworth}, \citenamefont {Davison}, \citenamefont {Rubin},\ and\
  \citenamefont {Salam}}]{Butterworth:2008iy}%
  \BibitemOpen
  \bibfield  {author} {\bibinfo {author} {\bibfnamefont {Jonathan~M.}\
  \bibnamefont {Butterworth}}, \bibinfo {author} {\bibfnamefont {Adam~R.}\
  \bibnamefont {Davison}}, \bibinfo {author} {\bibfnamefont {Mathieu}\
  \bibnamefont {Rubin}}, \ and\ \bibinfo {author} {\bibfnamefont {Gavin~P.}\
  \bibnamefont {Salam}},\ }\bibfield  {title} {\enquote {\bibinfo {title} {{Jet
  substructure as a new Higgs search channel at the LHC}},}\ }\href {\doibase
  10.1103/PhysRevLett.100.242001} {\bibfield  {journal} {\bibinfo  {journal}
  {Phys.Rev.Lett.}\ }\textbf {\bibinfo {volume} {100}},\ \bibinfo {pages}
  {242001} (\bibinfo {year} {2008})},\ \Eprint {http://arxiv.org/abs/0802.2470}
  {arXiv:0802.2470 [hep-ph]} \BibitemShut {NoStop}%
\bibitem [{\citenamefont {Krohn}\ \emph {et~al.}(2010)\citenamefont {Krohn},
  \citenamefont {Thaler},\ and\ \citenamefont {Wang}}]{Krohn:2009th}%
  \BibitemOpen
  \bibfield  {author} {\bibinfo {author} {\bibfnamefont {David}\ \bibnamefont
  {Krohn}}, \bibinfo {author} {\bibfnamefont {Jesse}\ \bibnamefont {Thaler}}, \
  and\ \bibinfo {author} {\bibfnamefont {Lian-Tao}\ \bibnamefont {Wang}},\
  }\bibfield  {title} {\enquote {\bibinfo {title} {{Jet Trimming}},}\ }\href
  {\doibase 10.1007/JHEP02(2010)084} {\bibfield  {journal} {\bibinfo  {journal}
  {JHEP}\ }\textbf {\bibinfo {volume} {1002}},\ \bibinfo {pages} {084}
  (\bibinfo {year} {2010})},\ \Eprint {http://arxiv.org/abs/0912.1342}
  {arXiv:0912.1342 [hep-ph]} \BibitemShut {NoStop}%
\bibitem [{\citenamefont {Ellis}\ \emph {et~al.}(2009)\citenamefont {Ellis},
  \citenamefont {Vermilion},\ and\ \citenamefont {Walsh}}]{Ellis:2009su}%
  \BibitemOpen
  \bibfield  {author} {\bibinfo {author} {\bibfnamefont {Stephen~D.}\
  \bibnamefont {Ellis}}, \bibinfo {author} {\bibfnamefont {Christopher~K.}\
  \bibnamefont {Vermilion}}, \ and\ \bibinfo {author} {\bibfnamefont
  {Jonathan~R.}\ \bibnamefont {Walsh}},\ }\bibfield  {title} {\enquote
  {\bibinfo {title} {{Techniques for improved heavy particle searches with jet
  substructure}},}\ }\href {\doibase 10.1103/PhysRevD.80.051501} {\bibfield
  {journal} {\bibinfo  {journal} {Phys.Rev.}\ }\textbf {\bibinfo {volume}
  {D80}},\ \bibinfo {pages} {051501} (\bibinfo {year} {2009})},\ \Eprint
  {http://arxiv.org/abs/0903.5081} {arXiv:0903.5081 [hep-ph]} \BibitemShut
  {NoStop}%
\bibitem [{\citenamefont {Ellis}\ \emph {et~al.}(2010)\citenamefont {Ellis},
  \citenamefont {Vermilion},\ and\ \citenamefont {Walsh}}]{Ellis:2009me}%
  \BibitemOpen
  \bibfield  {author} {\bibinfo {author} {\bibfnamefont {Stephen~D.}\
  \bibnamefont {Ellis}}, \bibinfo {author} {\bibfnamefont {Christopher~K.}\
  \bibnamefont {Vermilion}}, \ and\ \bibinfo {author} {\bibfnamefont
  {Jonathan~R.}\ \bibnamefont {Walsh}},\ }\bibfield  {title} {\enquote
  {\bibinfo {title} {{Recombination Algorithms and Jet Substructure: Pruning as
  a Tool for Heavy Particle Searches}},}\ }\href {\doibase
  10.1103/PhysRevD.81.094023} {\bibfield  {journal} {\bibinfo  {journal}
  {Phys.Rev.}\ }\textbf {\bibinfo {volume} {D81}},\ \bibinfo {pages} {094023}
  (\bibinfo {year} {2010})},\ \Eprint {http://arxiv.org/abs/0912.0033}
  {arXiv:0912.0033 [hep-ph]} \BibitemShut {NoStop}%
\bibitem [{\citenamefont {Dasgupta}\ \emph {et~al.}(2013)\citenamefont
  {Dasgupta}, \citenamefont {Fregoso}, \citenamefont {Marzani},\ and\
  \citenamefont {Salam}}]{Dasgupta:2013ihk}%
  \BibitemOpen
  \bibfield  {author} {\bibinfo {author} {\bibfnamefont {Mrinal}\ \bibnamefont
  {Dasgupta}}, \bibinfo {author} {\bibfnamefont {Alessandro}\ \bibnamefont
  {Fregoso}}, \bibinfo {author} {\bibfnamefont {Simone}\ \bibnamefont
  {Marzani}}, \ and\ \bibinfo {author} {\bibfnamefont {Gavin~P.}\ \bibnamefont
  {Salam}},\ }\bibfield  {title} {\enquote {\bibinfo {title} {{Towards an
  understanding of jet substructure}},}\ }\href {\doibase
  10.1007/JHEP09(2013)029} {\bibfield  {journal} {\bibinfo  {journal} {JHEP}\
  }\textbf {\bibinfo {volume} {1309}},\ \bibinfo {pages} {029} (\bibinfo {year}
  {2013})},\ \Eprint {http://arxiv.org/abs/1307.0007} {arXiv:1307.0007
  [hep-ph]} \BibitemShut {NoStop}%
\bibitem [{\citenamefont {Larkoski}\ \emph {et~al.}(2015)\citenamefont
  {Larkoski}, \citenamefont {Marzani},\ and\ \citenamefont
  {Thaler}}]{Larkoski:2015lea}%
  \BibitemOpen
  \bibfield  {author} {\bibinfo {author} {\bibfnamefont {Andrew~J.}\
  \bibnamefont {Larkoski}}, \bibinfo {author} {\bibfnamefont {Simone}\
  \bibnamefont {Marzani}}, \ and\ \bibinfo {author} {\bibfnamefont {Jesse}\
  \bibnamefont {Thaler}},\ }\bibfield  {title} {\enquote {\bibinfo {title}
  {{Sudakov Safety in Perturbative QCD}},}\ }\href {\doibase
  10.1103/PhysRevD.91.111501} {\bibfield  {journal} {\bibinfo  {journal} {Phys.
  Rev.}\ }\textbf {\bibinfo {volume} {D91}},\ \bibinfo {pages} {111501}
  (\bibinfo {year} {2015})},\ \Eprint {http://arxiv.org/abs/1502.01719}
  {arXiv:1502.01719 [hep-ph]} \BibitemShut {NoStop}%
\bibitem [{\citenamefont {Altarelli}\ and\ \citenamefont
  {Parisi}(1977)}]{Altarelli:1977zs}%
  \BibitemOpen
  \bibfield  {author} {\bibinfo {author} {\bibfnamefont {Guido}\ \bibnamefont
  {Altarelli}}\ and\ \bibinfo {author} {\bibfnamefont {G.}~\bibnamefont
  {Parisi}},\ }\bibfield  {title} {\enquote {\bibinfo {title} {{Asymptotic
  Freedom in Parton Language}},}\ }\href {\doibase
  10.1016/0550-3213(77)90384-4} {\bibfield  {journal} {\bibinfo  {journal}
  {Nucl. Phys.}\ }\textbf {\bibinfo {volume} {B126}},\ \bibinfo {pages} {298}
  (\bibinfo {year} {1977})}\BibitemShut {NoStop}%
\bibitem [{\citenamefont {Alwall}\ \emph {et~al.}(2014)\citenamefont {Alwall},
  \citenamefont {Frederix}, \citenamefont {Frixione}, \citenamefont {Hirschi},
  \citenamefont {Maltoni}, \citenamefont {Mattelaer}, \citenamefont {Shao},
  \citenamefont {Stelzer}, \citenamefont {Torrielli},\ and\ \citenamefont
  {Zaro}}]{Alwall:2014hca}%
  \BibitemOpen
  \bibfield  {author} {\bibinfo {author} {\bibfnamefont {J.}~\bibnamefont
  {Alwall}}, \bibinfo {author} {\bibfnamefont {R.}~\bibnamefont {Frederix}},
  \bibinfo {author} {\bibfnamefont {S.}~\bibnamefont {Frixione}}, \bibinfo
  {author} {\bibfnamefont {V.}~\bibnamefont {Hirschi}}, \bibinfo {author}
  {\bibfnamefont {F.}~\bibnamefont {Maltoni}}, \bibinfo {author} {\bibfnamefont
  {O.}~\bibnamefont {Mattelaer}}, \bibinfo {author} {\bibfnamefont {H.~S.}\
  \bibnamefont {Shao}}, \bibinfo {author} {\bibfnamefont {T.}~\bibnamefont
  {Stelzer}}, \bibinfo {author} {\bibfnamefont {P.}~\bibnamefont {Torrielli}},
  \ and\ \bibinfo {author} {\bibfnamefont {M.}~\bibnamefont {Zaro}},\
  }\bibfield  {title} {\enquote {\bibinfo {title} {{The automated computation
  of tree-level and next-to-leading order differential cross sections, and
  their matching to parton shower simulations}},}\ }\href {\doibase
  10.1007/JHEP07(2014)079} {\bibfield  {journal} {\bibinfo  {journal} {JHEP}\
  }\textbf {\bibinfo {volume} {07}},\ \bibinfo {pages} {079} (\bibinfo {year}
  {2014})},\ \Eprint {http://arxiv.org/abs/1405.0301} {arXiv:1405.0301
  [hep-ph]} \BibitemShut {NoStop}%
\bibitem [{\citenamefont {Sjostrand}\ \emph {et~al.}(2006)\citenamefont
  {Sjostrand}, \citenamefont {Mrenna},\ and\ \citenamefont
  {Skands}}]{Sjostrand:2006za}%
  \BibitemOpen
  \bibfield  {author} {\bibinfo {author} {\bibfnamefont {Torbjorn}\
  \bibnamefont {Sjostrand}}, \bibinfo {author} {\bibfnamefont {Stephen}\
  \bibnamefont {Mrenna}}, \ and\ \bibinfo {author} {\bibfnamefont {Peter~Z.}\
  \bibnamefont {Skands}},\ }\bibfield  {title} {\enquote {\bibinfo {title}
  {{PYTHIA 6.4 Physics and Manual}},}\ }\href {\doibase
  10.1088/1126-6708/2006/05/026} {\bibfield  {journal} {\bibinfo  {journal}
  {JHEP}\ }\textbf {\bibinfo {volume} {0605}},\ \bibinfo {pages} {026}
  (\bibinfo {year} {2006})},\ \Eprint {http://arxiv.org/abs/hep-ph/0603175}
  {arXiv:hep-ph/0603175 [hep-ph]} \BibitemShut {NoStop}%
\bibitem [{\citenamefont {Sjostrand}\ \emph {et~al.}(2008)\citenamefont
  {Sjostrand}, \citenamefont {Mrenna},\ and\ \citenamefont
  {Skands}}]{Sjostrand:2007gs}%
  \BibitemOpen
  \bibfield  {author} {\bibinfo {author} {\bibfnamefont {Torbjorn}\
  \bibnamefont {Sjostrand}}, \bibinfo {author} {\bibfnamefont {Stephen}\
  \bibnamefont {Mrenna}}, \ and\ \bibinfo {author} {\bibfnamefont {Peter~Z.}\
  \bibnamefont {Skands}},\ }\bibfield  {title} {\enquote {\bibinfo {title} {{A
  Brief Introduction to PYTHIA 8.1}},}\ }\href {\doibase
  10.1016/j.cpc.2008.01.036} {\bibfield  {journal} {\bibinfo  {journal}
  {Comput.Phys.Commun.}\ }\textbf {\bibinfo {volume} {178}},\ \bibinfo {pages}
  {852--867} (\bibinfo {year} {2008})},\ \Eprint
  {http://arxiv.org/abs/0710.3820} {arXiv:0710.3820 [hep-ph]} \BibitemShut
  {NoStop}%
\bibitem [{\citenamefont {Norrbin}\ and\ \citenamefont
  {Sjostrand}(2001)}]{Norrbin:2000uu}%
  \BibitemOpen
  \bibfield  {author} {\bibinfo {author} {\bibfnamefont {E.}~\bibnamefont
  {Norrbin}}\ and\ \bibinfo {author} {\bibfnamefont {T.}~\bibnamefont
  {Sjostrand}},\ }\bibfield  {title} {\enquote {\bibinfo {title} {{QCD
  radiation off heavy particles}},}\ }\href {\doibase
  10.1016/S0550-3213(01)00099-2} {\bibfield  {journal} {\bibinfo  {journal}
  {Nucl. Phys.}\ }\textbf {\bibinfo {volume} {B603}},\ \bibinfo {pages}
  {297--342} (\bibinfo {year} {2001})},\ \Eprint
  {http://arxiv.org/abs/hep-ph/0010012} {arXiv:hep-ph/0010012 [hep-ph]}
  \BibitemShut {NoStop}%
\bibitem [{\citenamefont {Bahr}\ \emph {et~al.}(2008)\citenamefont {Bahr},
  \citenamefont {Gieseke}, \citenamefont {Gigg}, \citenamefont {Grellscheid},
  \citenamefont {Hamilton} \emph {et~al.}}]{Bahr:2008pv}%
  \BibitemOpen
  \bibfield  {author} {\bibinfo {author} {\bibfnamefont {M.}~\bibnamefont
  {Bahr}}, \bibinfo {author} {\bibfnamefont {S.}~\bibnamefont {Gieseke}},
  \bibinfo {author} {\bibfnamefont {M.A.}\ \bibnamefont {Gigg}}, \bibinfo
  {author} {\bibfnamefont {D.}~\bibnamefont {Grellscheid}}, \bibinfo {author}
  {\bibfnamefont {K.}~\bibnamefont {Hamilton}},  \emph {et~al.},\ }\bibfield
  {title} {\enquote {\bibinfo {title} {{Herwig++ Physics and Manual}},}\ }\href
  {\doibase 10.1140/epjc/s10052-008-0798-9} {\bibfield  {journal} {\bibinfo
  {journal} {Eur.Phys.J.}\ }\textbf {\bibinfo {volume} {C58}},\ \bibinfo
  {pages} {639--707} (\bibinfo {year} {2008})},\ \Eprint
  {http://arxiv.org/abs/0803.0883} {arXiv:0803.0883 [hep-ph]} \BibitemShut
  {NoStop}%
\bibitem [{\citenamefont {Gleisberg}\ \emph {et~al.}(2009)\citenamefont
  {Gleisberg}, \citenamefont {Hoeche}, \citenamefont {Krauss}, \citenamefont
  {Schonherr}, \citenamefont {Schumann}, \citenamefont {Siegert},\ and\
  \citenamefont {Winter}}]{Gleisberg:2008ta}%
  \BibitemOpen
  \bibfield  {author} {\bibinfo {author} {\bibfnamefont {T.}~\bibnamefont
  {Gleisberg}}, \bibinfo {author} {\bibfnamefont {Stefan.}\ \bibnamefont
  {Hoeche}}, \bibinfo {author} {\bibfnamefont {F.}~\bibnamefont {Krauss}},
  \bibinfo {author} {\bibfnamefont {M.}~\bibnamefont {Schonherr}}, \bibinfo
  {author} {\bibfnamefont {S.}~\bibnamefont {Schumann}}, \bibinfo {author}
  {\bibfnamefont {F.}~\bibnamefont {Siegert}}, \ and\ \bibinfo {author}
  {\bibfnamefont {J.}~\bibnamefont {Winter}},\ }\bibfield  {title} {\enquote
  {\bibinfo {title} {{Event generation with SHERPA 1.1}},}\ }\href {\doibase
  10.1088/1126-6708/2009/02/007} {\bibfield  {journal} {\bibinfo  {journal}
  {JHEP}\ }\textbf {\bibinfo {volume} {02}},\ \bibinfo {pages} {007} (\bibinfo
  {year} {2009})},\ \Eprint {http://arxiv.org/abs/0811.4622} {arXiv:0811.4622
  [hep-ph]} \BibitemShut {NoStop}%
\bibitem [{\citenamefont {Dokshitzer}\ \emph {et~al.}(1993)\citenamefont
  {Dokshitzer}, \citenamefont {Khoze}, \citenamefont {Orr},\ and\ \citenamefont
  {Stirling}}]{Dokshitzer:1992nh}%
  \BibitemOpen
  \bibfield  {author} {\bibinfo {author} {\bibfnamefont {Yuri~L.}\ \bibnamefont
  {Dokshitzer}}, \bibinfo {author} {\bibfnamefont {Valery~A.}\ \bibnamefont
  {Khoze}}, \bibinfo {author} {\bibfnamefont {Lynne~H.}\ \bibnamefont {Orr}}, \
  and\ \bibinfo {author} {\bibfnamefont {W.~James}\ \bibnamefont {Stirling}},\
  }\bibfield  {title} {\enquote {\bibinfo {title} {{Properties of soft
  radiation near $t \bar{t}$ and $W^{-} W^{-}$ threshold}},}\ }\href {\doibase
  10.1016/0550-3213(93)90029-O} {\bibfield  {journal} {\bibinfo  {journal}
  {Nucl. Phys.}\ }\textbf {\bibinfo {volume} {B403}},\ \bibinfo {pages}
  {65--100} (\bibinfo {year} {1993})},\ \Eprint
  {http://arxiv.org/abs/hep-ph/9302250} {arXiv:hep-ph/9302250 [hep-ph]}
  \BibitemShut {NoStop}%
\bibitem [{\citenamefont {Orr}\ \emph {et~al.}(1993)\citenamefont {Orr},
  \citenamefont {Dokshitzer}, \citenamefont {Khoze},\ and\ \citenamefont
  {Stirling}}]{Orr:1993kd}%
  \BibitemOpen
  \bibfield  {author} {\bibinfo {author} {\bibfnamefont {Lynne~H.}\
  \bibnamefont {Orr}}, \bibinfo {author} {\bibfnamefont {Yuri~L.}\ \bibnamefont
  {Dokshitzer}}, \bibinfo {author} {\bibfnamefont {Valery~A.}\ \bibnamefont
  {Khoze}}, \ and\ \bibinfo {author} {\bibfnamefont {W.~James}\ \bibnamefont
  {Stirling}},\ }\bibfield  {title} {\enquote {\bibinfo {title} {{Gluon
  radiation and top width effects}},}\ }in\ \href@noop {} {\emph {\bibinfo
  {booktitle} {{2nd International Workshop on Physics and Experiments with
  Linear e+ e- Colliders Waikoloa, Hawaii, April 26-30, 1993}}}}\ (\bibinfo
  {year} {1993})\ \Eprint {http://arxiv.org/abs/hep-ph/9307338}
  {arXiv:hep-ph/9307338 [hep-ph]} \BibitemShut {NoStop}%
\bibitem [{\citenamefont {Khoze}\ \emph {et~al.}(1994)\citenamefont {Khoze},
  \citenamefont {Ohnemus},\ and\ \citenamefont {Stirling}}]{Khoze:1993ij}%
  \BibitemOpen
  \bibfield  {author} {\bibinfo {author} {\bibfnamefont {Valery~A.}\
  \bibnamefont {Khoze}}, \bibinfo {author} {\bibfnamefont {J.}~\bibnamefont
  {Ohnemus}}, \ and\ \bibinfo {author} {\bibfnamefont {W.~James}\ \bibnamefont
  {Stirling}},\ }\bibfield  {title} {\enquote {\bibinfo {title} {{Soft gluon
  radiation in hadronic t anti-t production}},}\ }\href {\doibase
  10.1103/PhysRevD.49.1237} {\bibfield  {journal} {\bibinfo  {journal} {Phys.
  Rev.}\ }\textbf {\bibinfo {volume} {D49}},\ \bibinfo {pages} {1237--1245}
  (\bibinfo {year} {1994})},\ \Eprint {http://arxiv.org/abs/hep-ph/9308359}
  {arXiv:hep-ph/9308359 [hep-ph]} \BibitemShut {NoStop}%
\bibitem [{CMS(2014)}]{CMS:2014fya}%
  \BibitemOpen
  \href@noop {} {\emph {\bibinfo {title} {{Boosted Top Jet Tagging at CMS}}}},\
  \bibinfo {type} {Tech. Rep.}\ \bibinfo {number} {CMS-PAS-JME-13-007}\
  (\bibinfo {year} {2014})\BibitemShut {NoStop}%
\bibitem [{\citenamefont {Cacciari}\ \emph {et~al.}(2008)\citenamefont
  {Cacciari}, \citenamefont {Salam},\ and\ \citenamefont
  {Soyez}}]{Cacciari:2008gp}%
  \BibitemOpen
  \bibfield  {author} {\bibinfo {author} {\bibfnamefont {Matteo}\ \bibnamefont
  {Cacciari}}, \bibinfo {author} {\bibfnamefont {Gavin~P.}\ \bibnamefont
  {Salam}}, \ and\ \bibinfo {author} {\bibfnamefont {Gregory}\ \bibnamefont
  {Soyez}},\ }\bibfield  {title} {\enquote {\bibinfo {title} {{The Anti-k(t)
  jet clustering algorithm}},}\ }\href {\doibase 10.1088/1126-6708/2008/04/063}
  {\bibfield  {journal} {\bibinfo  {journal} {JHEP}\ }\textbf {\bibinfo
  {volume} {0804}},\ \bibinfo {pages} {063} (\bibinfo {year} {2008})},\ \Eprint
  {http://arxiv.org/abs/0802.1189} {arXiv:0802.1189 [hep-ph]} \BibitemShut
  {NoStop}%
\bibitem [{\citenamefont {Cacciari}\ \emph {et~al.}(2012)\citenamefont
  {Cacciari}, \citenamefont {Salam},\ and\ \citenamefont
  {Soyez}}]{Cacciari:2011ma}%
  \BibitemOpen
  \bibfield  {author} {\bibinfo {author} {\bibfnamefont {Matteo}\ \bibnamefont
  {Cacciari}}, \bibinfo {author} {\bibfnamefont {Gavin~P.}\ \bibnamefont
  {Salam}}, \ and\ \bibinfo {author} {\bibfnamefont {Gregory}\ \bibnamefont
  {Soyez}},\ }\bibfield  {title} {\enquote {\bibinfo {title} {{FastJet User
  Manual}},}\ }\href {\doibase 10.1140/epjc/s10052-012-1896-2} {\bibfield
  {journal} {\bibinfo  {journal} {Eur.Phys.J.}\ }\textbf {\bibinfo {volume}
  {C72}},\ \bibinfo {pages} {1896} (\bibinfo {year} {2012})},\ \Eprint
  {http://arxiv.org/abs/1111.6097} {arXiv:1111.6097 [hep-ph]} \BibitemShut
  {NoStop}%
\bibitem [{\citenamefont {Dokshitzer}\ \emph {et~al.}(1997)\citenamefont
  {Dokshitzer}, \citenamefont {Leder}, \citenamefont {Moretti},\ and\
  \citenamefont {Webber}}]{Dokshitzer:1997in}%
  \BibitemOpen
  \bibfield  {author} {\bibinfo {author} {\bibfnamefont {Yuri~L.}\ \bibnamefont
  {Dokshitzer}}, \bibinfo {author} {\bibfnamefont {G.D.}\ \bibnamefont
  {Leder}}, \bibinfo {author} {\bibfnamefont {S.}~\bibnamefont {Moretti}}, \
  and\ \bibinfo {author} {\bibfnamefont {B.R.}\ \bibnamefont {Webber}},\
  }\bibfield  {title} {\enquote {\bibinfo {title} {{Better jet clustering
  algorithms}},}\ }\href@noop {} {\bibfield  {journal} {\bibinfo  {journal}
  {JHEP}\ }\textbf {\bibinfo {volume} {9708}},\ \bibinfo {pages} {001}
  (\bibinfo {year} {1997})},\ \Eprint {http://arxiv.org/abs/hep-ph/9707323}
  {arXiv:hep-ph/9707323 [hep-ph]} \BibitemShut {NoStop}%
\bibitem [{\citenamefont {Larkoski}\ and\ \citenamefont
  {Thaler}(2014)}]{Larkoski:2014bia}%
  \BibitemOpen
  \bibfield  {author} {\bibinfo {author} {\bibfnamefont {Andrew~J.}\
  \bibnamefont {Larkoski}}\ and\ \bibinfo {author} {\bibfnamefont {Jesse}\
  \bibnamefont {Thaler}},\ }\bibfield  {title} {\enquote {\bibinfo {title}
  {{Aspects of jets at 100 TeV}},}\ }\href {\doibase
  10.1103/PhysRevD.90.034010} {\bibfield  {journal} {\bibinfo  {journal}
  {Phys.Rev.}\ }\textbf {\bibinfo {volume} {D90}},\ \bibinfo {pages} {034010}
  (\bibinfo {year} {2014})},\ \Eprint {http://arxiv.org/abs/1406.7011}
  {arXiv:1406.7011 [hep-ph]} \BibitemShut {NoStop}%
\bibitem [{\citenamefont {Czakon}\ \emph {et~al.}(2013)\citenamefont {Czakon},
  \citenamefont {Fiedler},\ and\ \citenamefont {Mitov}}]{Czakon:2013goa}%
  \BibitemOpen
  \bibfield  {author} {\bibinfo {author} {\bibfnamefont {Micha¿}\ \bibnamefont
  {Czakon}}, \bibinfo {author} {\bibfnamefont {Paul}\ \bibnamefont {Fiedler}},
  \ and\ \bibinfo {author} {\bibfnamefont {Alexander}\ \bibnamefont {Mitov}},\
  }\bibfield  {title} {\enquote {\bibinfo {title} {{Total Top-Quark
  Pair-Production Cross Section at Hadron Colliders Through
  $O(¿\frac{4}{S})$}},}\ }\href {\doibase 10.1103/PhysRevLett.110.252004}
  {\bibfield  {journal} {\bibinfo  {journal} {Phys. Rev. Lett.}\ }\textbf
  {\bibinfo {volume} {110}},\ \bibinfo {pages} {252004} (\bibinfo {year}
  {2013})},\ \Eprint {http://arxiv.org/abs/1303.6254} {arXiv:1303.6254
  [hep-ph]} \BibitemShut {NoStop}%
\bibitem [{\citenamefont {Chatrchyan}\ \emph {et~al.}(2013)\citenamefont
  {Chatrchyan} \emph {et~al.}}]{Chatrchyan:2012jua}%
  \BibitemOpen
  \bibfield  {author} {\bibinfo {author} {\bibfnamefont {Serguei}\ \bibnamefont
  {Chatrchyan}} \emph {et~al.} (\bibinfo {collaboration} {CMS}),\ }\bibfield
  {title} {\enquote {\bibinfo {title} {{Identification of b-quark jets with the
  CMS experiment}},}\ }\href {\doibase 10.1088/1748-0221/8/04/P04013}
  {\bibfield  {journal} {\bibinfo  {journal} {JINST}\ }\textbf {\bibinfo
  {volume} {8}},\ \bibinfo {pages} {P04013} (\bibinfo {year} {2013})},\ \Eprint
  {http://arxiv.org/abs/1211.4462} {arXiv:1211.4462 [hep-ex]} \BibitemShut
  {NoStop}%
\bibitem [{\citenamefont {Aad}\ \emph {et~al.}(2009)\citenamefont {Aad} \emph
  {et~al.}}]{Aad:2009wy}%
  \BibitemOpen
  \bibfield  {author} {\bibinfo {author} {\bibfnamefont {G.}~\bibnamefont
  {Aad}} \emph {et~al.} (\bibinfo {collaboration} {ATLAS}),\ }\bibfield
  {title} {\enquote {\bibinfo {title} {{Expected Performance of the ATLAS
  Experiment - Detector, Trigger and Physics}},}\ }\href@noop {} {\  (\bibinfo
  {year} {2009})},\ \Eprint {http://arxiv.org/abs/0901.0512} {arXiv:0901.0512
  [hep-ex]} \BibitemShut {NoStop}%
\bibitem [{\citenamefont {Bayatian}\ \emph {et~al.}(2006)\citenamefont
  {Bayatian} \emph {et~al.}}]{Bayatian:2006zz}%
  \BibitemOpen
  \bibfield  {author} {\bibinfo {author} {\bibfnamefont {G.L.}\ \bibnamefont
  {Bayatian}} \emph {et~al.} (\bibinfo {collaboration} {CMS}),\ }\href@noop {}
  {\emph {\bibinfo {title} {{CMS physics: Technical design report}}}},\
  \bibinfo {type} {Tech. Rep.}\ \bibinfo {number} {CERN-LHCC-2006-001,
  CMS-TDR-008-1}\ (\bibinfo {year} {2006})\BibitemShut {NoStop}%
\bibitem [{\citenamefont {Gallicchio}\ and\ \citenamefont
  {Schwartz}(2010)}]{Gallicchio:2010sw}%
  \BibitemOpen
  \bibfield  {author} {\bibinfo {author} {\bibfnamefont {Jason}\ \bibnamefont
  {Gallicchio}}\ and\ \bibinfo {author} {\bibfnamefont {Matthew~D.}\
  \bibnamefont {Schwartz}},\ }\bibfield  {title} {\enquote {\bibinfo {title}
  {{Seeing in Color: Jet Superstructure}},}\ }\href {\doibase
  10.1103/PhysRevLett.105.022001} {\bibfield  {journal} {\bibinfo  {journal}
  {Phys. Rev. Lett.}\ }\textbf {\bibinfo {volume} {105}},\ \bibinfo {pages}
  {022001} (\bibinfo {year} {2010})},\ \Eprint {http://arxiv.org/abs/1001.5027}
  {arXiv:1001.5027 [hep-ph]} \BibitemShut {NoStop}%
\bibitem [{\citenamefont {Abazov}\ \emph {et~al.}(2011)\citenamefont {Abazov}
  \emph {et~al.}}]{Abazov:2011vh}%
  \BibitemOpen
  \bibfield  {author} {\bibinfo {author} {\bibfnamefont {Victor~Mukhamedovich}\
  \bibnamefont {Abazov}} \emph {et~al.} (\bibinfo {collaboration} {D0}),\
  }\bibfield  {title} {\enquote {\bibinfo {title} {{Measurement of color flow
  in $\mathbf{t\bar{t}}$ events from $\mathbf{p\bar{p}}$ collisions at
  $\mathbf{\sqrt{s}=1.96}$ TeV}},}\ }\href {\doibase
  10.1103/PhysRevD.83.092002} {\bibfield  {journal} {\bibinfo  {journal} {Phys.
  Rev.}\ }\textbf {\bibinfo {volume} {D83}},\ \bibinfo {pages} {092002}
  (\bibinfo {year} {2011})},\ \Eprint {http://arxiv.org/abs/1101.0648}
  {arXiv:1101.0648 [hep-ex]} \BibitemShut {NoStop}%
\bibitem [{\citenamefont {Aad}\ \emph {et~al.}(2015)\citenamefont {Aad} \emph
  {et~al.}}]{Aad:2015lxa}%
  \BibitemOpen
  \bibfield  {author} {\bibinfo {author} {\bibfnamefont {Georges}\ \bibnamefont
  {Aad}} \emph {et~al.} (\bibinfo {collaboration} {ATLAS}),\ }\bibfield
  {title} {\enquote {\bibinfo {title} {{Measurement of colour flow with the jet
  pull angle in $t\bar{t}$ events using the ATLAS detector at $\sqrt{s}=8$
  TeV}},}\ }\href {\doibase 10.1016/j.physletb.2015.09.051} {\bibfield
  {journal} {\bibinfo  {journal} {Phys. Lett.}\ }\textbf {\bibinfo {volume}
  {B750}},\ \bibinfo {pages} {475--493} (\bibinfo {year} {2015})},\ \Eprint
  {http://arxiv.org/abs/1506.05629} {arXiv:1506.05629 [hep-ex]} \BibitemShut
  {NoStop}%
\bibitem [{\citenamefont {de~Favereau}\ \emph {et~al.}(2014)\citenamefont
  {de~Favereau}, \citenamefont {Delaere}, \citenamefont {Demin}, \citenamefont
  {Giammanco}, \citenamefont {Lema"tre}, \citenamefont {Mertens},\ and\
  \citenamefont {Selvaggi}}]{deFavereau:2013fsa}%
  \BibitemOpen
  \bibfield  {author} {\bibinfo {author} {\bibfnamefont {J.}~\bibnamefont
  {de~Favereau}}, \bibinfo {author} {\bibfnamefont {C.}~\bibnamefont
  {Delaere}}, \bibinfo {author} {\bibfnamefont {P.}~\bibnamefont {Demin}},
  \bibinfo {author} {\bibfnamefont {A.}~\bibnamefont {Giammanco}}, \bibinfo
  {author} {\bibfnamefont {V.}~\bibnamefont {Lema"tre}}, \bibinfo {author}
  {\bibfnamefont {A.}~\bibnamefont {Mertens}}, \ and\ \bibinfo {author}
  {\bibfnamefont {M.}~\bibnamefont {Selvaggi}} (\bibinfo {collaboration}
  {DELPHES 3}),\ }\bibfield  {title} {\enquote {\bibinfo {title} {{DELPHES 3, A
  modular framework for fast simulation of a generic collider experiment}},}\
  }\href {\doibase 10.1007/JHEP02(2014)057} {\bibfield  {journal} {\bibinfo
  {journal} {JHEP}\ }\textbf {\bibinfo {volume} {02}},\ \bibinfo {pages} {057}
  (\bibinfo {year} {2014})},\ \Eprint {http://arxiv.org/abs/1307.6346}
  {arXiv:1307.6346 [hep-ex]} \BibitemShut {NoStop}%
\end{thebibliography}%

\end{document}